\pdfoutput=1
\documentclass[10pt, conference, compsocconf]{IEEEtran}
\usepackage{cite}
\usepackage{adjustbox}
\usepackage{graphicx}
\usepackage{booktabs}
\usepackage{multirow}
\usepackage{multicol}
\usepackage{amsmath}
\usepackage{amssymb}
\usepackage{subcaption}
\usepackage{ifthen}
\usepackage{ifpdf}
\usepackage{color}
\usepackage{url}
\usepackage{array}
\usepackage{tabularx}
\let\labelindent\relax
\usepackage{enumitem}
\usepackage{comment}
\usepackage{algorithm}
\usepackage{algorithmicx}
\usepackage{caption}
\usepackage{lipsum}
\usepackage{siunitx}
\usepackage{fancyvrb}
\usepackage{listings}
\usepackage[bookmarks=false]{hyperref}

\makeatletter
\newcommand{\verbatimfont}[1]{\renewcommand{\verbatim@font}{\ttfamily#1}}
\makeatother

\newcommand\boldfont[1]{\textbf{#1}}
\usepackage{tgcursor}

\def\IEEEbibitemsep{0pt plus .5pt}

\definecolor{rd} {rgb} {1.0,0.0,0.0}
\definecolor{lg} {rgb} {0.7,0.3,0.9}
\definecolor{db} {rgb} {0.0,0.0,0.7}
\definecolor{dg} {rgb} {0.0,0.7,0.0}
\definecolor{dr} {rgb} {0.7,0.0,0.0}
\definecolor{gr} {rgb} {0.8,0.4,0.1}

\newcommand{\John } [1] {{{\color{db}[John] #1}}}
\newcommand{\YZH  } [1] {{{\color{dg}[YZH]  #1}}}
\newcommand{\YC   } [1] {{{\color{lg}[YC]#1}}}
\newcommand{\Yuduo} [1] {{{\color{gr}[Yuduo] #1}}}
\newcommand{\Carl } [1] {{{\color{dr}[Carl] #1}}}

\newcommand{\final} {0}

\ifthenelse{\equal{\final}{1}} {\renewcommand{\John }[1]{}}{}
\ifthenelse{\equal{\final}{1}} {\renewcommand{\YZH }[1]{}}{}
\ifthenelse{\equal{\final}{1}} {\renewcommand{\YC }[1]{}}{}
\ifthenelse{\equal{\final}{1}} {\renewcommand{\Yuduo }[1]{}}{}
\ifthenelse{\equal{\final}{1}} {\renewcommand{\Carl }[1]{}}{}

\providecommand{\shortcite}[1]{\cite{#1}}
\ifdefined\textln\relax\else\newcommand{\textln}[1]{{\fontfamily{pplx}\selectfont #1}}\fi
\begin{document}


%\title{Performance Characterization of High-Level Large-Scale GPU Graph
% Analytics Abstractions}
% YZH's try of the title: Towards a High-level Programming Model for
% Large-scale Graph Processing on the GPU
%\title{Towards a High-Level Programming Model for Large-Scale Graph
%Processing on the GPU}
\title{Multi-GPU Graph Analytics}

% \author{\IEEEauthorblockN{Michael Shell}
% \IEEEauthorblockA{School of Electrical and\\Computer Engineering\\
% Georgia Institute of Technology\\
% Atlanta, Georgia 30332--0250\\
% Email: http://www.michaelshell.org/contact.html}
% \and
% \IEEEauthorblockN{Homer Simpson}
% \IEEEauthorblockA{Twentieth Century Fox\\
% Email: homer@thesimpsons.com}
% \and
% \IEEEauthorblockN{James Kirk\\ and Montgomery Scott}
% \IEEEauthorblockA{Starfleet Academy\\
% San Francisco, California 96678--2391\\
% Telephone: (800) 555--1212\\
\author{
\IEEEauthorblockN{Yuechao Pan, Yangzihao Wang, Yuduo Wu, Carl Yang, and John D.
Owens}
\IEEEauthorblockA{University of California, Davis\\
Email: \{ychpan, yzhwang, yudwu, ctcyang, jowens\}@ucdavis.edu}
}

\maketitle

\begin{abstract}

  \label{sec:abstract}
  We present a single-node, multi-GPU programmable graph processing
library that allows programmers to easily extend single-GPU graph
algorithms to achieve scalable performance on large graphs with
billions of edges. Directly using the single-GPU implementations,
our design only requires programmers to specify a few
algorithm-dependent concerns, hiding most multi-GPU related
implementation details. We analyze the theoretical and practical
limits to scalability in the context of varying graph primitives and
datasets. We describe several optimizations, such as direction
optimizing traversal, and a just-enough memory allocation scheme,
for better performance and smaller memory consumption. Compared
to previous work, we achieve best-of-class performance across
operations and datasets, including excellent strong and weak
scalability on most primitives as we increase the number of
GPUs in the system.

\end{abstract}

\begin{IEEEkeywords}
GPU; multi GPU; parallel graph processing;
\end{IEEEkeywords}

\section{Introduction}
\label{sec:intro}
The potential advantages in performance, performance per dollar, and
performance per watt of the modern graphics processing unit (GPU) over
the traditional CPU~\cite{Keckler:2011:GAT} has led to a recent focus
on GPU graph analytics~\cite{Zhong:2014:MSG, Fu:2014:MAH,
Khorasani:2014:CVG, Wang:2016:GAH:nourl}. However, scalable GPU graph
analytics frameworks today---those beyond one GPU---are still
primarily in the research domain. In general, today's GPU graph
analytics frameworks, which we summarize in
Section~\ref{sec:litsurvey}, do not deliver both high performance and
scalability while maintaining programmability and algorithm
generality.

Single-GPU (``1GPU'') frameworks deliver excellent performance
on graphs that fit into the GPU's (limited) memory. Scaling to larger
graphs and/or achieving higher performance require new approaches. We
see three directions for scalability: multiple GPUs on a node
(``mGPU''); the most common approach to date, multiple nodes
(``mNode''); or leveraging the storage of a larger CPU memory
(``out-of-core''). These directions are non-exclusive; future
scalable systems may use more than one. Our performance results motivate
our belief that mGPU graph processing should become the fundamental
building block of GPU graph analytics.

Our work builds on our open-source ``Gunrock''~\cite{Wang:2016:GAH:nourl}
graph-processing library for GPUs, whose programming model we
summarize in Section~\ref{sec:gunrock}. While Gunrock previously
targeted 1GPU and graphs that fit into 1GPU's memory,
we optimize and extend it in this work to mGPU\@.
We believe that the conclusions we make in this paper will apply
to other GPU graph frameworks as well.

To achieve high performance, scalability,
programmability, and generality, we address several key questions:
\begin{itemize}
  \item What is a general mGPU graph processing model?
  \item How to transform 1GPU programs to support mGPU\@?
  \item What data should be communicated, when, and how?
  \item How do we synchronize GPUs during computation?
  \item What is the indicator for convergence?
  \item What are the potential limiting factors to scalability?
  \item What are the optimizations for these limiting factors?
\end{itemize}

Addressing all of these goals is challenging. Supporting
programmability and generality tends toward a high-level, flexible
framework that abstracts away low-level details. However, performance
and scalability concerns instead suggest low-level implementations
that can efficiently leverage the underlying hardware. We also note
other factors that can potentially limit performance and scalability:
the type of graph partitioner used, the topology of the underlying
graphs, and the necessary synchronization and communication patterns
of each individual primitive.

Our work makes the following contributions:
\begin{enumerate}
\item Our mGPU graph processing library meets the above goals.
  Our framework allows programmers to easily extend
  1GPU primitives to utilize mGPU's capabilities.
\item We perform a detailed experimental analysis on potential
  limiting factors to scalability. We identify communication
  bandwidth; synchronization latency; efficient use of GPU memory; and
  partitioning strategy as the most significant obstacles. For partitioning
  strategy, we conclude that minimizing the size of \emph{partition
    borders} as opposed to the traditional partitioners' target of
  minimizing \emph{edge cuts} is the right strategy for our system.
\item We design and implement generalized optimizations that
  effectively target these limiting factors, enhancing our performance
  and scalability. Our novel optimizations include efficient mGPU
  direction-optimizing traversal, and a just-enough memory allocation
  strategy that makes efficient use of GPU memory.
\item We achieve best-in-class performance on mGPU graph
  primitives, outperforming primitive-specific implementations on
  similar machine configurations. On 6 GPUs, we achieve more than 900 GTEPS
  (billion edges traversed per seconds)
  peak performance for direction-optimizing breath-first search
  (DOBFS)~\cite{Beamer:2012:DBS}, and 2.63$\times$, 2.57$\times$,
  2.00$\times$, 1.96$\times$, 3.86$\times$ geometric mean speedups
  as compared to 1GPU, and over various datasets for
  breadth-first search (BFS), single-source shortest path (SSSP),
  connected components (CC), betweenness centrality (BC) and PageRank
  (PR) respectively.
\end{enumerate}

\section{Related Work}
\label{sec:background}
\subsection{Scalable GPU Graph Libraries}
\label{sec:litsurvey}

Numerous frameworks have targeted scalable graph analytics with
multi-GPU approaches. We argue that in general, no previous multi-GPU
work achieves our balance of high performance with programmability.
%\YZH{should these two multi-GPUs be mGPU instead? Unless you are talking
%about both mGPU and mNode, which seems to be the case.}
%\YC{Yes, both mGPU and mNode, so still multi-GPU here.}

Merrill et al.~\cite{Merrill:2012:SGG} presented the first notable
linear parallelization of the BFS algorithm on the GPU\@. Their 1GPU
and mGPU implementations achieve excellent performance. In their mGPU
implementation, vertices are distributed to GPUs, data related to
remote vertices are fetched via peer memory access. Their approach
only targets BFS, and adversely affects programmability by forcing
programmers to handle cross-GPU data access within main computing steps. The
peer memory access limits hardware compatibility, and also introduces
load imbalance when accessing both local and remote vertices, which
reduces performance.

The parallel BFS work by Fu et al.~\cite{Fu:2014:MAH} extends the
expand-contract BFS algorithm by Merrill et al.\ to GPU clusters. They
propose a 2D partitioning method, and use MPI to contract columns on
the edge frontiers after each expand step. The communication pattern
limits data access within 1 hop, and thus restricts algorithm
generality. Large edge frontiers transmitted between GPUs cause large
communication overheads and limit scalability.

Bisson, Bernaschi, and Mastrofano~\cite{Bisson:2015:PDB} focused on
building an mNode BFS implementation. They also utilize
a 2D-partitioning scheme to reduce the amount of communication
required. However, because of their use of costly atomic operations,
their performance is limited.

Enterprise~\cite{Liu:2015:EBG} is an mGPU work that targets BFS using
BFS-specific optimizations. Their work achieves excellent performance
on rmat graphs, but lacks the generality to target algorithms beyond
BFS\@.

McLaughlin and Bader~\cite{McLaughlin:2014:SAH} targeted BC on GPU
clusters, which distributed BFS work for different source vertices to
different nodes. Its performance scales well in large part due to its
novel use of task parallelism, but a task-parallel strategy is not
applicable to most graph algorithms. Their framework also duplicates
the graph across GPUs, limiting its scalability to graphs that can fit
on 1GPU\@.

Medusa~\cite{Zhong:2014:MSG} was the pioneering mGPU graph library,
taking a more general approach. It partitions the graph using
Metis~\cite{Karypis:1998:FHQ}, makes replications for neighbor
vertices within $n$ hops, and updates vertex-associated values every
$n$ iterations. Their framework is limited in algorithm generality,
because it cannot express algorithms that jump beyond the $n$-hop
limit, such as Soman et al.'s CC algorithm~\cite{Soman:2010:AFG}.
Compared to its successors, it does not achieve top performance. And
due to the data replication caused by a large number of vertices
within $n$ hops of a partition boundary, their framework is not
scalable in memory usage.

Totem~\cite{Gharaibeh:2014:ELS} is a graph processing engine for
GPU-CPU hybrid systems. It either processes the workload on the CPU or
transmits it to the GPU according to a performance estimation model.
This approach has the potential to solve the long-tail problem on
GPUs, and overcome GPU memory size limitations. However, it has
limitations in algorithm generality, because it can only work with
algorithms that only access direct neighbors. Repeatedly moving
data between CPUs and GPUs is costly, which makes
scalability an issue.

Daga et al.~\cite{Daga:2014:EBF} explored using an accelerated processing unit
(APU, a single-chip CPU+GPU heterogeneous processor) to overcome
the PCIe bandwidth limitation, but the APU's memory bandwidth is significantly smaller
than a discrete GPU, which hampers overall performance.

GraphReduce~\cite{Sengupta:2015:GPL} is an out-of-core graph processing
library for GPU\@. It uses a Gather-Apply-Scatter (GAS) framework, so
it inherits GAS's programmability and algorithm generality. Its
out-of-core approach addresses the challenge of the GPU's limited
memory. However, it must stream the graph to the GPU during the
computation, making the PCIe bus a performance bottleneck. Its use of
only 1GPU also makes it unable to achieve performance
scalability.

Frog~\cite{Shi:2015:OAG,Shi:2015:FAG} differs from other frameworks
here in requiring (expensive) preprocessing to color the graph
into sets of independent vertices. With the colored graph, they can
process colors asynchronously. However, performance is restricted by
visiting all edges in each single iteration.
%A faster
%coloring computation would be an interesting technology that might
%allow asynchronous execution to become a part of our framework.

More recently, Groute~\cite{Ben-Nun:2017:GAA} leveraged asynchronous
computation to demonstrate impressive multi-GPU performance particularly
on high-diameter, road-network-like graphs, and primitives that can benefit
from prioritized data communication, such as SSSP and CC.\footnote{The Groute work
  was published after this paper completed peer review; we compare
  Gunrock's performance against Groute on the Gunrock website.
  \url{http://gunrock.github.io/gunrock/doc/latest/md_stats_groute.html}}

\subsection{Gunrock: GPU Graph Analytics}
\label{sec:gunrock}

Our GPU-based graph analytics framework, Gunrock, targets both
programmability and performance, and achieved them on
1GPU~\cite{Wang:2016:GAH:nourl}. It proposes a data-centric programming model that
presents graph primitives as a series of parallel graph operations on frontiers,
which is a group of vertices or edges that are actively participating in the
computation. It currently supports three ways to manipulate the frontier:
\begin{description}
\item[Advance] generates a new frontier by visiting the neighbors of
  the current frontier;
\item[Filter] generates a new frontier by selecting a subset of the
  current frontier based on programmer-specified criteria;
\item[Computation] executes an operation on all elements in the
  current frontier. This can be combined for efficiency with advance or
  filter.
\end{description}
Gunrock programs define graph algorithms as a sequence of the above
three steps, beginning with an initial frontier and running to
convergence.
%(e.g., an empty frontier, or a maximum number of iterations).

\section{Our MGPU Graph Programming Abstraction}
\label{sec:framework}
Our mGPU programming model is designed to balance programming
complexity and performance. As much as possible, our philosophy is to
enable programmers to specify algorithms at a high level, targeting
a 1GPU implementation, while allowing our underlying mGPU framework
to manage the necessary parallelization and communication details.
Thus we make the following design decisions:
\begin{enumerate}

\item An mGPU implementation in our system uses a 1GPU
  primitive without modification; all mGPU machinery is transparent to
  it. This isolation not only simplifies the 1GPU to mGPU
  transformation, but also allows optimizations, either to primitives
  or to the underlying framework, to apply to both 1- and
  m-GPU cases.

\item To support mGPU primitives, programmers must specify the
  information listed in Section~\ref{sec:1GPU_to_mGPU}.

\item Both because of GPU memory limitations and to leverage inter-GPU
  parallelism, our system partitions graphs across GPUs. We do not restrict the selection of the partitioner, leaving that decision to the programmer
  (Section~\ref{sec:partitioner}).

\end{enumerate}

\subsection{Terminology}
\label{sec:terminology}

We define a graph $G(V, E)$ by its vertices $V$ and edges $E$, and its
diameter as $D$. When partitioned, the $i^{th}$ GPU only holds a
subgraph of $G$, denoted as $G_i(V_i, E_i)$, where $V_i$ and $E_i$ are
the vertices and the edges stored on it. $B_{i,j}$ is the outgoing
vertex border from GPU $i$ to GPU $j$, and $B_i$ is the union of all
$B_{i,j}$, including duplications. We use $L_i$ to represent all
vertices hosted by GPU $i$; $L_i$ may be a subset of $V_i$, because
$V_i$ also contains remote proxy vertices (more
in Section~\ref{sec:communication-strategy}). $n$ is the number of
GPUs.

Because our system is bulk-synchronous across all GPUs, we use the BSP
model~\cite{Valiant:1990:ABM} as a useful tool to analyze
performance-limiting factors (Section~\ref{sec:analysis}). In the BSP model,
$W$ is the cost of local computation on a single node (analyzed in
Section~\ref{sec:algorithms}); $H$ is the number of messages transmitted
(communication volume, the size of data transmitted between GPUs);
$g$ is the time to deliver a single message
under continuous traffic conditions (we use the inverse of inter-GPU
communication bandwidth); $S$ is the number of supersteps (iterations);
and $l$ is the synchronization cost (per-iteration
overhead). We use $C$ to represent the cost of communication computation,
which is the computation required to facilitate inter-GPU communications.

\subsection{Extending 1GPU Programs to mGPU}
\label{sec:1GPU_to_mGPU}

\begin{figure}
  \centering
  \includegraphics[width=0.45\textwidth]{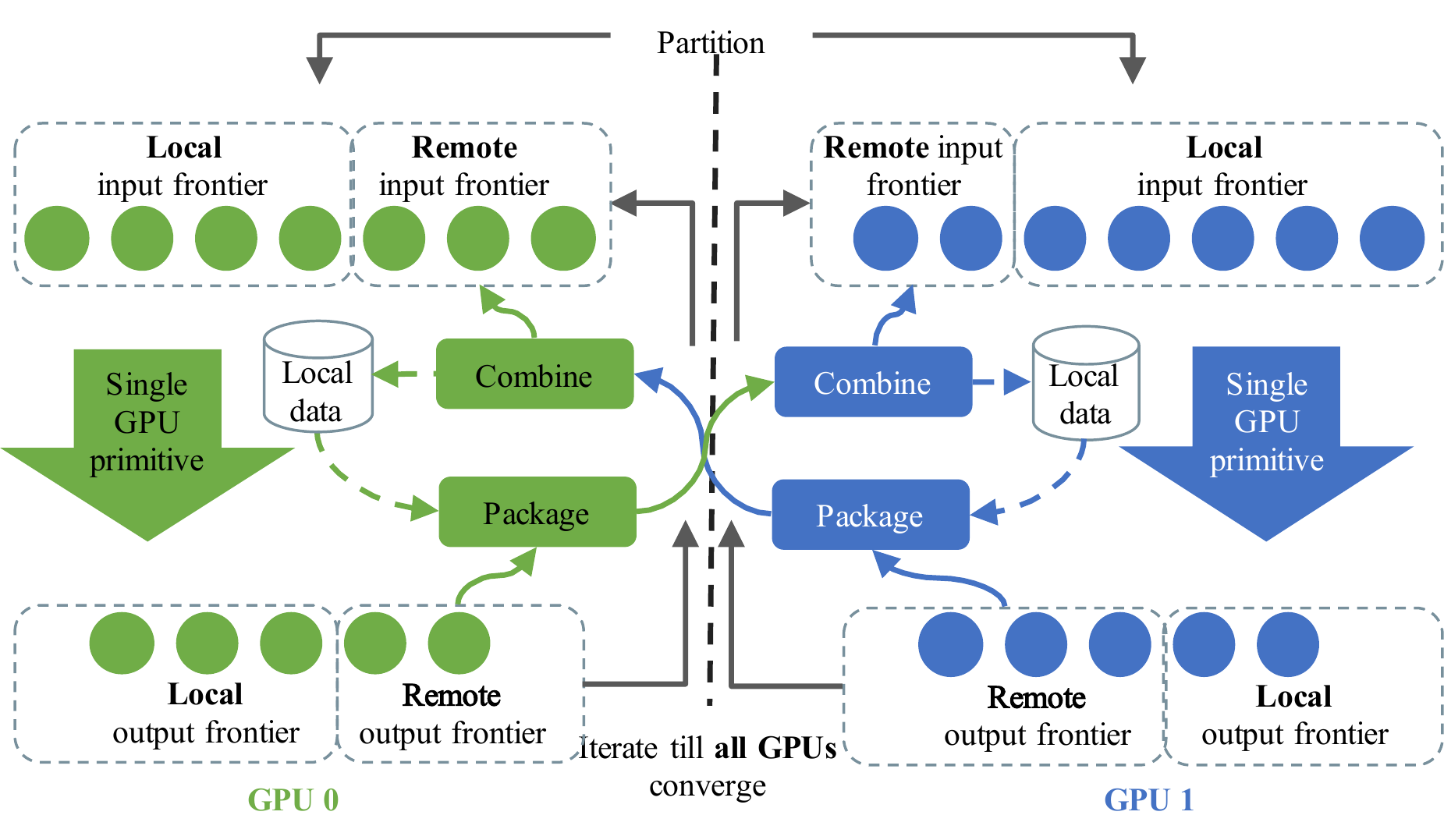}
  \centering
  \caption{mGPU Framework highlighting communications
     \label{fig:framework}}
\end{figure}

Our mGPU framework is illustrated in Fig.~\ref{fig:framework}. The
core of an mGPU primitive is an unmodified 1GPU primitive, which
we extend to mGPUs by using the iteration synchronization
point to exchange data between GPUs. By making our mGPU framework
transparent to 1GPU primitives, we separate the concerns of per-GPU
computation and inter-GPU communication: as long as the input
frontiers for each iteration are prepared correctly, and the
per-vertex associative values are updated properly before they are
used in the next iteration, a core graph primitive does not need to
know whether a vertex is hosted on a local or remote
GPU\@.

At the end of an iteration, a 1GPU primitive concludes its
computation and globally synchronizes before beginning the next
iteration. At that point, our framework takes over, performing the
following steps: it splits the output frontier of vertices into
local and remote sub-frontiers, then packages the remote sub-frontier
with its primitive-specific associated data, identified by the
programmers, such as labels or predecessor vertex ids, and pushes
the packaged data to peer GPUs. When a GPU receives data,
our framework combines that data with that
GPU's local data, and (if necessary) also adds the received vertices
into its input frontier for the next iteration. A GPU
that hosts vertex $V$ may receive updates to $V$ from multiple remote
GPUs. It must combine those updates into a single value for $V$. The
programmer specifies how that combining must be done---for instance,
taking the minimum value from all updates, as in BFS or SSSP---but the
framework will actually perform the combining.

Our mGPU framework also initializes the computation by
partitioning the graph and its associated per-\{vertex, edge\} data,
reordering or relabeling if necessary, distributing all data to
the correct GPUs, and initializing the starting frontier.

Programmers must specify the following information:

\begin{description}
\item[Core single-GPU primitive] Use Gunrock operators to define a
  series of operations on input frontiers.
\item[Data to communicate] What kinds of data associated with vertices
  must be pushed to remote GPUs?
  We have not seen primitives that require per-edge
  communication between GPUs, and argue that any such primitive will
  scale poorly based on the large volume and computation workload
  required by per-edge communication.
\item[Combining remote and local data] Specify the operation to
  combine local and (possibly multiple) received data at the
  beginning of an iteration, except the first one.
\item[Stop condition] Define the local and/or global stop condition so
  that each local GPU will properly exit its computing iteration when
  the algorithm finishes.
\end{description}

\noindent
and the framework handles all other aspects:

\begin{description}
\item[Split frontier] Split the output frontier of an iteration into
  local and remote sub-frontiers.
\item[Package data] Package the remote sub-frontiers with the associated
  data that specified by the programmer. Data packaging can be done together
  with frontier splitting.
\item[Push to remote GPUs] Manage communication so that each GPU
  pushes the right information to the right GPU for use in the next
  iteration.
\item[Merge local and received sub-frontiers] Using the combiner specified by
  the programmer, efficiently merge the local sub-frontier with all received
  sub-frontiers to get the input frontier for the next iteration.
\item[Manage GPUs] Our framework manages each GPU by a dedicated CPU thread
  to avoid false dependencies between GPUs. It also uses multiple GPU
  streams on a GPU to overlap computation and communication, by separating
  them into different streams. We synchronize and establish dependencies
  between GPUs without CPU intervention by
  using \texttt{cudaStreamWaitEvent()}.

%   \John{How do
%     we overlap computation with communication here? Seems to me that
%     we do local computation, then compute what to send, then stop
%     computing while we send it.}
%   \YC{For most cases, that's true, and computation of next iteration can only
% start after data from all peer GPUs arrived and combined. For some primitives,
% computation of the next iteration can start with only partial input frontier,
% and no need to wait. This is a bit complicated, and I will try not to cover it
% in this paper}
\end{description}

\subsection{Vertex Duplication and Communication Strategy}
\label{sec:communication-strategy}

We partition the graph (Section~\ref{sec:partitioner}) as a
pre-processing stage, and currently support
partitioners that do edge cuts, i.e., vertices are distributed to
individual GPUs, together with their outgoing edges. To isolate the
computation to local data only, remote vertices need to be duplicated
locally. We implemented two strategies for this duplication:

\begin{description}
\item[Duplicate-1-hop:] create a local proxy vertex only for the
  immediate remote neighbors of $L_i$ on GPU$_i$; vertices in $V_i$ are
  renumbered with continuous IDs.

\item[Duplicate-all:] create a local proxy vertex for \emph{every}
  remote vertex, i.e., force $V_i$ to be $V$. We
  still distribute $E$, so remote vertices in $V_i$ have 0 outgoing
  edges on GPU $i$.
\end{description}

We also implemented two strategies for communication:

\begin{description}
\item[Broadcast:] in each iteration, each GPU broadcasts the
  whole generated frontier to all other GPUs.

\item[Selective-communicate:] we send frontier vertices to
  only their hosting GPUs or to the GPUs that host their proxies.
  This requires a splitting step on the vertex frontier to assemble a
  separate sub-frontier of vertices to each remote GPU\@.
\end{description}

The programmer can choose the strategies.
Duplicate-1-hop uses less memory space, but requires ID conversion for
communication; on the other hand, duplicate-all requires no ID
conversion but uses more memory. Broadcasting saves the work required
to split the frontier, but consumes more memory and communication
bandwidth, and introduces a higher computation workload when combining
received data. Selective communication requires less memory and
communication bandwidth, but it cannot bypass the splitting step. If
an algorithm only needs to access the immediate neighbors of incoming
or outgoing edges, then duplicate-1-hop and selective-communication
are better choices; otherwise, algorithms that access both incoming
and outgoing neighbors (e.g., DOBFS), or that visit vertices with more
than one hop distance (e.g., CC), require broadcasting, and usually
use the duplicate-all strategy.

\section{Algorithms in Our Abstraction}
\label{sec:algorithms}
\begin{table*}[t]
  \centering
    \begin{tabular}{lcccc}
        \toprule
            Primitive & Computation ($W$) & Communication Computation ($C$) & Communication Volume ($H$) & Iterations ($S$)\\
        \midrule
            BFS   & $O(|E_i|)$           & $O(|V_i|)$
                  & $O(|B_i|)$           & $\sim D / 2$ \\
            DOBFS & $O(a  \times |E_i|)$   & $O(|V|)$
                  & $O((n-1)\times |V|)$ & $\sim D / 2$\\
            SSSP  & $O(b  \times |E_i|)$  & $O(b\times|V_i|)$
                  & $O(2b \times |B_i|)$ & $\sim b\times D / 2$ \\
            BC    & $O(2  \times |E_i|)$  & $O(2\times |V_i| + |V|)$
                  & $O(5|B_i| + 2(n-1) \times |L_i|)$ & $\sim D / 2$ \\
            CC    & $\log(D/2) \times O(|E_i|)$ & $S\times O(|V_i|)$
                  & $S\times O(2|V_i|)$  & 2--5\\
            PR    & $S\times O(|E_i|)$   & $S\times O(|B_i|)$
                  & $S\times O(|B_i|)$   & data-dependent     \\
        \bottomrule
    \end{tabular}
    \caption {Summary of Algorithms.
      %Computation is the computation
      %complexity of the primitive; communication computation is the
      %computation required to facilitate inter-GPU communications;
      %communication volume is the size of data transmitted between
      %GPUs, and $S$ is the number of iterations to complete the
      %computation.
      Terminology is summarized in
      Section~\ref{sec:terminology}.\label{tab:algorithms}}
\end{table*}

%\DeclareCaptionFormat{algor}{%
%  \hrulefill\par\offinterlineskip\vskip1pt%
%    \textbf{#1#2}#3\offinterlineskip\hrulefill}
%\DeclareCaptionStyle{algori}{singlelinecheck=off,format=algor,labelsep=space}
%\captionsetup[algorithm]{style=algori}

\makeatletter
\newenvironment{breakablealgorithm}
  {% \begin{breakablealgorithm}
   \begin{center}
     \refstepcounter{algorithm}% New algorithm
     \hrule height.8pt depth0pt \kern2pt % \@fs@pre for \@fs@ruled
     \renewcommand{\caption}[2][\relax]{% Make a new \caption
       {\raggedright\textbf{\ALG@name~\thealgorithm} ##2\par}%
       \ifx\relax##1\relax % #1 is \relax
         \addcontentsline{loa}{algorithm}{\protect\numberline{\thealgorithm}##2}%
       \else % #1 is not \relax
         \addcontentsline{loa}{algorithm}{\protect\numberline{\thealgorithm}##1}%
       \fi
       \kern2pt\hrule\kern2pt
     }
  }{% \end{breakablealgorithm}
     \kern2pt\hrule\relax% \@fs@post for \@fs@ruled
    \end{center}
  }
\makeatother

We implemented six graph primitives with our framework, several of
which are straightforward extensions (from the programmer's perspective)
from 1GPU implementations. Included in Appendix~\ref{app:bfs} is
an example BFS implementation using the mGPU framework, with programmer
provided blocks highlighted. Because DOBFS has different runtime and
communication properties, and thus different scaling behavior, we
consider it as a separate algorithm in our scalability analysis.
SSSP and BC have similar scaling properties to BFS, while CC and PR
are both non-traversal primitives that operate on all vertices and all
edges of the graph. We thus select DOBFS, BFS, and PR as
representative primitives for further analysis,
and summarize all six algorithms in Table~\ref{tab:algorithms}.
Except for DOBFS and for graphs that have too little computation
to fill the GPU on an iteration, most primitives are bounded by computation.

\begin{breakablealgorithm}%[h]
\caption{Multi-GPU BFS}
\begin{algorithmic}[1]
\Statex\noindent\textbf{Vertex duplication}: Duplicate-all. We trade
memory usage for better performance for BFS.

\Statex\textbf{Computation}: An advance kernel followed by a filter
kernel, as introduced by Merrill et al.~\cite{Merrill:2012:SGG}. $W{\in}O(|E_i|)$.

\Statex\textbf{Communication}: Selective-communicate. Only the remote
vertices are sent.

\Statex\textbf{Combination}: If a received vertex has not been visited
before, update its label and place it in the input frontier on
the next iteration. $H{\in}O(|B_i|)$, and $C{\in}O(|V_i|)$.

\Statex\textbf{Convergence}: All frontiers are empty. $S{\approx}D / 2$.
\label{algo:bfs}
\end{algorithmic}
\end{breakablealgorithm}

\begin{breakablealgorithm}%[H]
\caption{Multi-GPU DOBFS}
\begin{algorithmic}[2]
\Statex\noindent\textbf{Vertex duplication}: Duplicate-all. It couples
better with the broadcast communication strategy.

\Statex\textbf{Computation}: Summarized in
Section~\ref{sec:dobfs-optimization}. For graphs with high average
out-degrees, $W{\in}O(|L_i|)$; for
other graphs in practice, $W{\in}O(a\times|E_i|)$ where
$a < 1$.

\Statex\textbf{Communication}: Broadcast, because an upcoming iteration may use
either the forward or backward direction.

\Statex\textbf{Combination}: Same as BFS\@.
$H{\in}O(|V|)$ and $C{\in}O((n-1)|V|)$.

\Statex\textbf{Convergence}: Same as BFS\@. $S{\approx}D / 2$.
\label{algo:dobfs}
\end{algorithmic}
\end{breakablealgorithm}

\begin{breakablealgorithm}%[H]
\caption{Multi-GPU PR}
\begin{algorithmic}[3]
\Statex\noindent\textbf{Vertex duplication}: Either duplicate-all
or duplicate-1-hop. The remote sub-frontiers do not change over
iterations. We get all these sub-frontiers during the
initialization step, and only send ranking values during actual
computation. There is no significant performance or memory usage
difference between these two, and we use duplicate-all to better
trace the program.

\Statex\textbf{Computation}: A filter kernel updating the PR values
(except 1$^{st}$ iteration), followed by an advance kernel
accumulating the PR values for each vertex. $W{\in}O(|E_i|)$.

\Statex\textbf{Communication}: Selective-communicate. Push
locally accumulated ranks of each vertex to its hosting GPU.

\Statex\textbf{Combination}: Do an \texttt{atomicAdd}
to combine received rank with the local copy. $H{\in}O(|B_i|)$
and $C{\in}O(|B_i|)$.

\Statex\textbf{Convergence}: Terminates when all ranking value updates
are smaller than a pre-defined threshold ratio, or a given maximum
number of iterations is reached. $S$ does not
affect the scalability.

\label{algo:pr}
\end{algorithmic}
\end{breakablealgorithm}

%%% Local Variables:
%%% mode: latex
%%% TeX-master: "../main"
%%% End:

%\section{Implementation}
%\label{sec:abstraction}
%\input{tex/abstraction}
\section{Performance Limiting Factor Analysis}
\label{sec:analysis}
In this section we describe and analyze potential performance
bottlenecks for our mGPU implementation that are specific to graph
computation and mGPU systems. The BSP computation model~\cite{Valiant:1990:ABM}
states that the total computation cost of a parallel program can be
expressed as $W + Hg + Sl$.

\subsection{Communication}

The inter-GPU bandwidth $1 / g$ is determined by the system. Certainly
any mGPU system should
enable the highest-bandwidth connection. For instance,
enabling peer GPU-GPU communication on our system (K40s on
the same PCIe3 root hub) increases GPU-GPU bandwidth from
$\sim$16~GB/s to $\sim$20~GB/s with a corresponding latency decrease
from $\sim$25~us to $\sim$7.5~us.

% \begin{figure}
%   \centering
%   \includegraphics[width=\columnwidth]{fig/Bandwidth.pdf}
%   \centering
%   \caption{Slowdown of primitives as a function of artificially
%     increased communication volume (an $x$-axis value of $x$ means $x$
%     times more communication than our actual implementation). DOBFS is
%     impacted more than other primitives; larger vertex counts are
%     affected more than small ones.\label{fig:bandwidth}
%     }
% \end{figure}

What software \emph{can} affect is the communication volume $H$.
$H$ is perhaps the most important factor in scaling for a given primitive.
We list $H$ for individual primitives in Table~\ref{tab:algorithms}.
To get a clearer
idea how $H$ affects the performance, we artificially increased it
and found that in general, runtime varies linearly with the increase of $H$.
We also found that increasing $H$ affects DOBFS more than BFS and PR,
because $W$ and $H$ of DOBFS are close in scale,
especially when running on rmat graphs (both are roughly in $O(|V|)$),
whereas $W$ is larger than $H$ for BFS and PR\@.
Datasets with high vertex counts suffered more from $H$ increases.
Another possible factor that may influence performance is
communication latency, but it's only a small portion of $l$,
and even when we artificially increase
it by a factor of ten, we see no appreciable difference in performance.

\subsection{Synchronization}
\label{sec:synchronization}

% \begin{figure}
%   \centering
%   \includegraphics[width=\columnwidth]{fig/Overhead.pdf}
%   \centering
%   \caption{Runtime of BFS when per iteration work load is minimal, for
%     calculation of iteration overheads. LB is the normal BFS primitive, and
%     KF is the one with kernel fusion, which is introduced in
% Section~\ref{sec:fusion}. The
%     numbers in legends indicate the number of GPUs used.
%     \label{fig:overhead}}
% \end{figure}

The per-iteration synchronization latency $l$ includes the effects of
kernel launch overheads ($\sim$3~\si{\us} per kernel) during
primitive computation, load imbalance
between GPUs, and API and kernel launching overheads of the
communication-computation kernels. In our experiments, $l$
is significant when the other parts run in the
sub-\si{\us} range or $S$ is large.

The GPU also needs a large workload to maintain high processing
rates~\cite{Wu:2015:PCF:nourl}. If per-GPU workloads aren't large
enough, kernel launch overheads also occupy a large portion of the
total running time. Traversal on road networks is one example that
suffers from both launch latency and GPU under-utilization; one
iteration of even a large road network traversal doesn't have enough
work to keep even 1GPU busy. We also see this overhead
when processing, for instance, DOBFS
on rmat, where many iterations only take a few \si{\us}.

To study the effect of $l$, we let each GPU visit
only 1 vertex and 1 edge in each iteration. This is the smallest
per-iteration workload possible, and the BFS running time on it can be
used as an measurement of $l$. The running time is linear with $S$,
and the average per-iteration time for large $S$ for \{1,2,3,4\} GPUs
is \{66.8, 124, 142, 188\}~\si{\us}. The jump from one to two GPUs
reflects the effect of inter-GPU synchronization and communication
latency.

\subsection{Partitioner}
\label{sec:partitioner}

\begin{figure}
  \centering
  \includegraphics[width=\columnwidth]{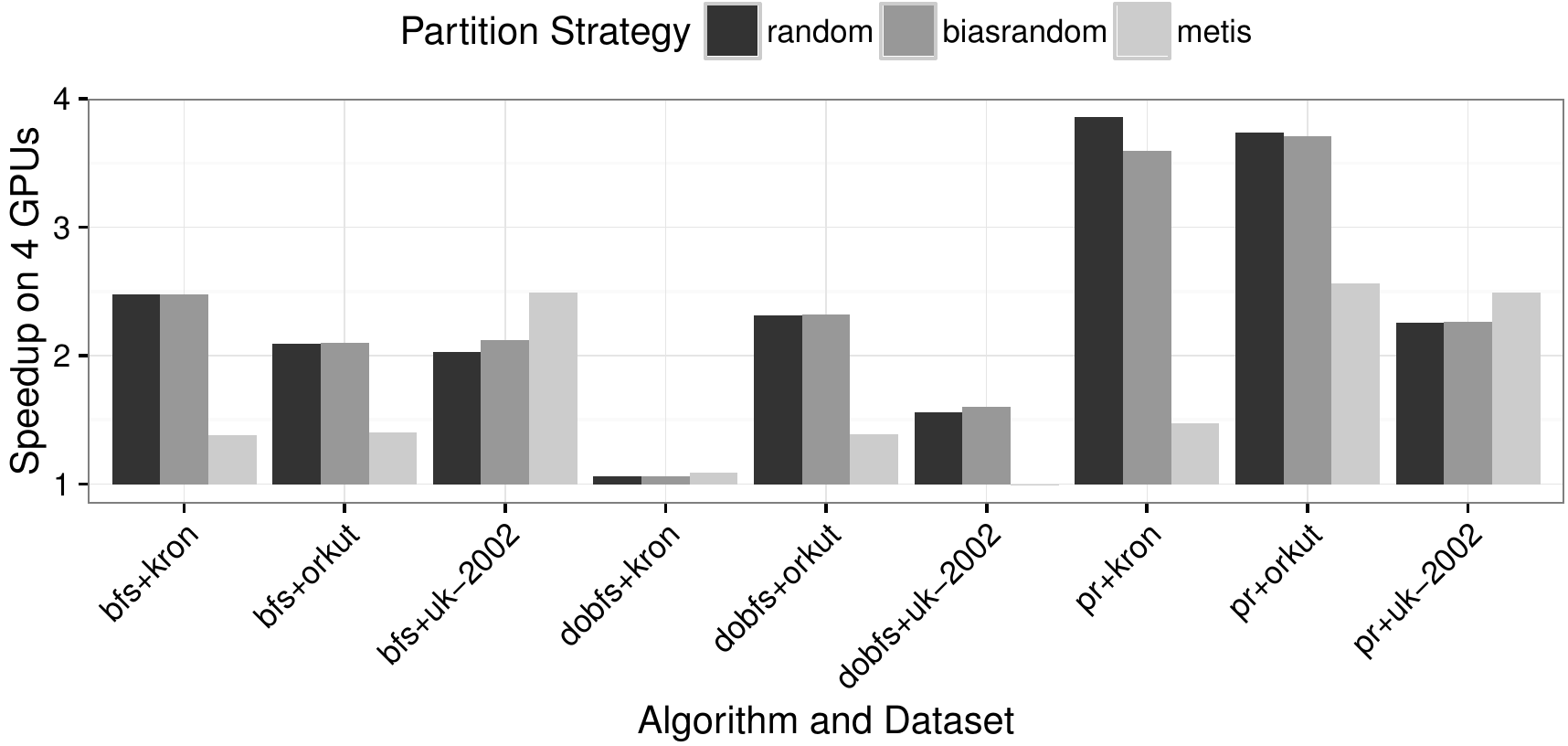}
  \centering
  \caption{Performance impact of partitioners on 3 primitives
  $\times$ 3 datasets.\label{fig:partitioner}}
\end{figure}

We recognize that good partitioners can help increase mGPU graph
processing performance. Most partitioners attempt to minimize the
number of edges cut across partitions. However, in our system, it is
instead the size of partition borders ($B_i$, the number of vertices
on partition edges, as summarized in Table~\ref{tab:algorithms}) that
is most important to our performance. This is because our framework
communicates values associated with vertices, and multiple cut edges
from the same GPU that point to the same remote vertex only need to
transmit one set of values regarding that vertex.

To gain insight into partitioner behavior, we use three different
partitioners, listed in increasing order of partitioner runtime:
\emph{random} (randomly assign vertices to GPUs), \emph{biased random}
(like random, but biased toward assigning a vertex to a GPU that
contains more of its neighbors), and
\emph{Metis}~\cite{Karypis:1998:FHQ}. We summarize their effects in
Fig.~\ref{fig:partitioner}. While the random partitioner captures no
graph locality, it does achieve excellent load balancing, and performs
fairly well across our tests. Biased random tries to reduce the border
size without affecting the load balancing too much, and shows performance
very close to the random partitioner. Metis only wins in a few situations,
with small margins, but takes a much longer time to partition. With
this in mind, all other experiments in this paper use the random
partitioner. Without ideal partitioner candidates, we chose to make
our partitioner interface modular and allow users to specify any
existing partitioner or implement their own; we ensure that the
framework and primitives will run correctly regardless of the
choice of partitioner.

\section{Optimizations}
\label{sec:optimization}
With 1GPU Gunrock, we begin with a framework that already has numerous
optimizations for good performance. We add several more
to specifically address mGPU operation and the performance-limiting
factors from Section~\ref{sec:analysis}.

\subsection{Direction-optimizing Traversal}
\label{sec:dobfs-optimization}
Traditional BFS performs a forward (``push'') traversal where vertices
in the current frontier add their unvisited neighbors to the output
frontier. Beamer et al.'s DOBFS~\cite{Beamer:2012:DBS} adds
the ability to perform a backward
(``pull'') traversal, beginning from a frontier of all unvisited
vertices, to visit parent vertices. If one of those parent
vertices is in the current frontier, visiting all other edges can be
skipped. This ``edge skipping'' can significantly improve BFS
performance for small-diameter graphs, but mapping existing
implementations to a distributed context is a challenge.

Beamer et al.\ implemented this operation by scanning all vertices
and processing the unvisited ones. This is inefficient,
and introduces load imbalance between visited and unvisited vertices.
Our previous 1GPU implementation had two deficiencies when
parallelized across GPUs.

First, our 1GPU advance kernel parallelizes across edges and thus
cannot efficiently skip edges once a parent is found. We added an advance mode
that parallelizes across vertices, thus serializing edge visits and
allowing us to stop work when we discover a valid parent. We then split
the unvisited vertex frontier from the previous iteration into two
parts, newly-discovered vertices and unvisited vertices. The newly
discovered vertex frontier is important for our mGPU implementation,
because it gives a direction-independent view for the framework (advances
in both directions output the newly discovered vertices),
and also a cost-free transformation from backward to forward.

Second, the traditional DOBFS computation for switching between push
and pull would require additional computation (potentially of the same scale
of the actual traversal) to get the number
of edges needed to visit in the next iteration. We change the
direction-selection condition to only require inputs that are already
available. Let $Q$ be the current frontier and $U$ and $P$ the
unvisited and visited vertices. We can estimate the number of forward
edges visited as $FV = \frac{|Q|\cdot|E_i|}{|V_i|}$ and the number of
backward edges visited as $BV = \frac{|U|\cdot|V_i|}{|P|}$. We begin
with forward traversal; then at the beginning of each subsequent
iteration, if we see that $FV > BV \cdot do\_a$, we switch from
forward to backward; if $FV < BV \cdot do\_b$, we switch from
backward to forward. Because every time we switch from forward to
backward, we must scan all vertices for unvisited ones, we only allow
this switch once. The optimal values of $do\_a$ and $do\_b$ for
similar graph types appear to be consistent; for example,
$do\_a = 0.01$ and $do\_b = 0.1$ gives good performance for social
graphs. We found these parameters are mostly mGPU-independent,
i.e., the same set of parameters can be used for different numbers of
GPUs.

These optimizations permit a significantly more efficient mGPU
DOBFS, one that outperforms previous BFS and DOBFS implementations by
a significant margin, but also uncovers a more fundamental bottleneck.
DOBFS's principal computation advantage is effectively reducing $W$
to $O(a\cdot|E_i|)$, where $a$ is less than 1.
For graphs where $|E_i|{\gg}|V_i|$, such as rmat
graphs with large edge factors, $W$ reduces to $O(|V_i|)$.
But because the upcoming iteration may use either direction, which
essentially requires sending the newly discovered frontiers to all peers
(i.e., broadcast), $H$ and $C$ thus increase
to $O(|V|)$, which can be on par with $W$. The
result is an implementation that is primarily bound by communication
with flat strong and weak scaling behavior
(Section~\ref{sec:dobfs-results}). Reducing communication cost is the
priority for future mGPU DOBFS implementations.

%\begin{figure}

%  \centering

%  \includegraphics[width=0.48\textwidth]{fig/dummy.png}

%  \centering

%  \caption{Overlap of computation and communication}
%  \label{fig:overlap}
%\end{figure}
%A particular GPU timeline showing overlaping computation and communication, can
%get from nvprofiler. %Fig.~\ref{fig:overlap}

\subsection{Just-enough Memory Allocation}
\label{sec:careful-memory-management}

\begin{figure}
  \centering
  \includegraphics[width=\columnwidth]{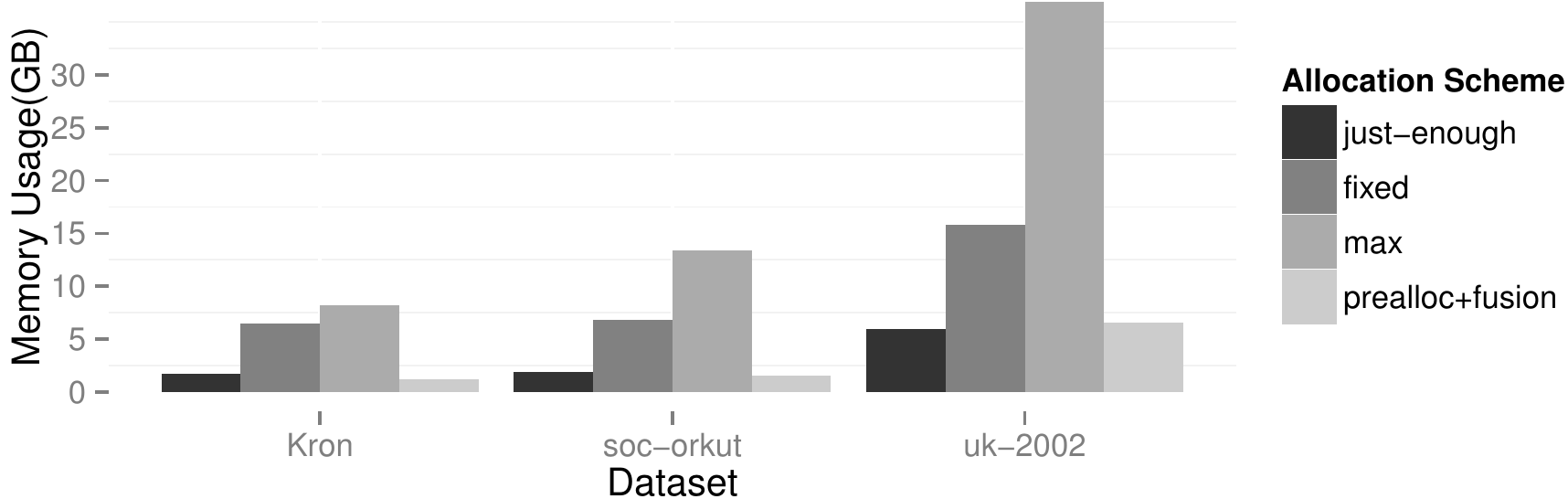}
  \centering
  \caption{Memory consumption using different allocation schemes on
    kron, soc-orkut, and uk-2002 running BFS\@. Details of the schemes
    are described in Section~\ref{sec:careful-memory-management}.
\label{fig:memory} }
\end{figure}

Because GPU memory capacity is limited, it is crucial to use it
efficiently, particularly for large graphs. What makes this
challenging is that iterative graph primitives usually produce
frontiers with a size that is unknown until the finish of an advance
or filter kernel. One option is to allocate memory that is large
enough to handle any case, e.g., a size $|E|$ array for advance.
However, this maximum allocation artificially limits the size of the
subgraph we can place onto one GPU, which either (a) requires us to
use more GPUs to solve a particular problem or (b) limits our
scalability to less than we could potentially achieve.

Instead of worst-case allocation, we implement a \emph{just-enough}
memory allocation scheme to use our memory more efficiently. We make a
reasonable estimate of memory allocation before computation and then
reallocate if this allocation is insufficient. In practice, our
``reasonable'' estimates are usually sufficient so reallocation, which
is expensive, is infrequent. We mainly reallocate before advance,
filter, or communicate operations. For advance, we leverage Gunrock's
load-balancing computations to correctly compute output size or add an
extra reduction to determine output size. For filter, the output size
is at most the size of input, and for most filters, the output size is
capped by $|V_i|$. For communication, the required size is given by
the framework.

We compare just-enough memory allocation against 3 alternatives
(Fig.~\ref{fig:memory}): a fixed preallocation using sizing factors
calculated from previous runs of similar graphs; a maximum allocation;
and preallocation
plus kernel fusion (in Section~\ref{sec:kernel-fusion}). The
just-enough memory allocation is still in effect when these
alternatives are used, to prevent illegal memory
access, although this only happens rarely. Each of the different
memory allocation schemes have near-identical computation times.
Just-enough allocation is
critical in reducing our memory footprint, which allows us to fit
larger subgraphs into memory. Consequently, we can achieve higher
performance with fewer GPUs than other frameworks that lack
sophisticated memory management strategies
(Section~\ref{sec:compare}). Our implementation of (DO)BFS, SSSP and
BC use preallocation plus kernel fusion
(Section~\ref{sec:kernel-fusion}); we use fixed preallocation for CC
and PR, as their memory requirements can be determined before running
these primitives.

%\subsection{Other optimizations}
%\label{sec:other_optimizations}

%\paragraph{Idempotence}

%BFS by Merrill et al.~\cite{Merrill:2012:SGG} uses idempotent operations
%to eliminate atomic operations, instead using
%a single-bit bitmask to indicate whether a vertex has been visited.
%Our work extends this idea to mGPU by setting the bitmask for vertices
%in received sub-frontiers.

\subsection{Kernel Fusion}
\label{sec:kernel-fusion}

Kernel fusion (automatically combining two sequential kernels into
one) is a well-known technique for high-performance GPU graph
analytics~\cite{Merrill:2012:SGG,Wu:2015:PCF:nourl}. Our
previous 1GPU work~\cite{Wang:2016:GAH:nourl} fused Gunrock compute operators with advance or
filter operators. In this work we added the
ability to fuse an advance operator with a filter operator that
follows it. In addition to the usual advantages of reducing kernel
launch overhead and increasing producer-consumer locality, this
particular fusing eliminates the need to store the intermediate
frontier (potentially as large as $O(|E|)$) in GPU memory,
enabling us to store larger subgraphs per GPU.

% \begin{figure}
%  \centering
%  \includegraphics[width=\columnwidth]{fig/Optimization.pdf}
%  \centering
%  \caption{Performance impact of optimizations. KF\@: kernel fusion, ID\@:
%  idempotence, DO\@: direction optimization.\label{fig:optimization}}
% \end{figure}

\section{Results}
\label{sec:results}
We begin by summarizing the results that we present in more detail later in this
section.

\begin{itemize}
\item Primitives in our framework scale reasonably well from 1 to 6
  GPUs (geometric mean of speedup: 2.52$\times$ across five primitives),
  except for DOBFS, whose scalability is limited by communication
  overhead.
\item (DO)BFS and PR show good weak scaling. BFS
  and PR exhibit strong scaling, but DOBFS does not.
\item We compared our performance against
  previously-published in-core multi-GPU systems on the datasets
  highlighted by those systems. In general, we
  significantly outperform other systems given the same number of GPUs, and often
  systems with many more GPUs.
\end{itemize}

\subsection{Experimental Setup}

\begin{table*}[t]
  \centering
  \resizebox{\linewidth}{!}{
    \centering
    \begin{tabular}{@{}c c r r r | c c r r r | c c r r r@{}}
      \toprule
      group & name & \multicolumn{1}{c}{$|V|$}
                   & \multicolumn{1}{c}{$|E|$}
                   & \multicolumn{1}{c}{$D$} &
      group & name & \multicolumn{1}{c}{$|V|$}
                   & \multicolumn{1}{c}{$|E|$}
                   & \multicolumn{1}{c}{$D$} &
      group & name & \multicolumn{1}{c}{$|V|$}
                   & \multicolumn{1}{c}{$|E|$}
                   & \multicolumn{1}{c}{$D$}\\
      \midrule
      soc  & soc-LiveJournal1 & 4.85M & 85.7M & 13 &
      web  & indochina-2004   & 7.41M & 302M  & 24 &
      rmat & rmat\_n20\_512   & 1.05M & 728M  & 6.26$^*$\\

      soc  & hollywood-2009   & 1.14M & 113M  & 8 &
      web  & uk-2002          & 18.5M & 524M  & 25 &
      rmat & rmat\_n21\_256   & 2.10M & 839M  & 7.22$^*$\\

      soc  & soc-orkut        & 3.00M & 213M  & 7 &
      web  & arabic-2005      & 22.7M & 1.11B & 28 &
      rmat & rmat\_n22\_128   & 4.19M & 925M  & 7.56$^*$\\

      soc  & soc-sinaweibo    & 58.7M & 523M  & 5 &
      web  & uk-2005          & 39.5M & 1.57B & 23 &
      rmat & rmat\_n23\_64    & 8.39M & 985M  & 8.32$^*$\\

      soc  & soc-twitter-2010 & 21.3M & 530M  & 15 &
      web  & webbase-2001     & 118M  & 1.71B & 379 &
      rmat & rmat\_n24\_32    & 16.8M & 1.02B & 8.61$^*$\\

           &                  &       &       & &
           &                  &       &       & &
      rmat & rmat\_n25\_16    & 33.6M & 1.05B & 9.06$^*$\\

      \bottomrule
    \end{tabular}
  }
  \centering
  \caption{Datasets we used to evaluate our work. $|V|$ and $|E|$ are
    vertex and edge counts; $d$ is the graph diameter, $^*$ indicates 
    an approximated diameter
    computed by multiple run of random-sourced BFS\@.\label{tab:dataset-info}}
\end{table*}

We run most tests on nodes with 6 NVIDIA Tesla K40 cards, a 10-core
Intel Xeon E5-2690 v2, and 128~GB CPU memory, running on CentOS 6.6
with CUDA 7.5 (both driver and runtime) and gcc 4.8.4. We conduct
strong and weak scaling experiments on 2 systems: (1) 4 NVIDIA Tesla
K80 cards (each with 2 GPUs and 12~GB DRAM/GPU) and (2) 4 Tesla P100s
(PCIe, 16~GB DRAM, CUDA 8.0). Direct peer-to-peer inter-GPU communication is
enabled in groups of 4 GPUs where appropriate. All programs are
compiled with the -O3 flag and set to target the actual streaming
multiprocessor generation of the GPU hardware.

The dataset information is listed in Table~\ref{tab:dataset-info} in
three representative groups. The real-world graphs are from the UF
sparse matrix collection~\cite{Davis:1994:UOF} and the Network Data
Repository~\cite{Rossi:2015:NDR}. The ``soc'' and ``web'' groups are
online social networks and web crawls of different domains. For SSSP,
edge values are randomly generated integers from [0, 64].
We implement a GPU-based R-MAT ``rmat'' graph
generator faithful to GTgraph~\cite{Bader:2006:GAS}; the rmat
parameters are \{A, B, C, D\} = \{0.57, 0.19, 0.19, 0.05\}. All three
kinds of graphs follow a power-law distribution. Road networks, and
high-diameter, low-degree graphs in general, have very different
scalability characteristics than power-law graphs. They have
insufficient parallelism to saturate even 1GPU, much less
mGPUs; as a result, iteration overhead occupies a significant
portion of the runtime, and we observed performance decreases on
mGPU\@. If otherwise unspecified, all graphs we use are
converted to undirected graphs. Self-loops and duplicated edges are removed.
All tests have been repeated at
least 10 times with average runtime used for results. Computations are
verified for correctness. The instructions and the scripts to reproduce
the results can be found at \href{https://github.com/gunrock/gunrock/tree/master/dataset/test-scripts/ipdps17}{\url{https://github.com/gunrock/gunrock/tree/master/dataset/test-scripts/ipdps17}}.

% Results for BFS, SSSP, BC and CC are verified against Boost
% 1.58~\cite{2013:BCL} on a CPU\@. For PR, the CPU reference
% implementation takes too long ($>$ 1 hour) to converge, so we
% verify our GPU implementation up to medium datasets with CPU
% implementation. For large datasets, we verify that the sum of ranking
% values are always close to 1, when PR runs in normalized mode with
% compensation (ranking of 0-outdegree vertices are evenly
% distributed to all vertices in the graph).

\subsection{Overall Results}

\begin{figure}
  \centering
  \includegraphics[width=\columnwidth]{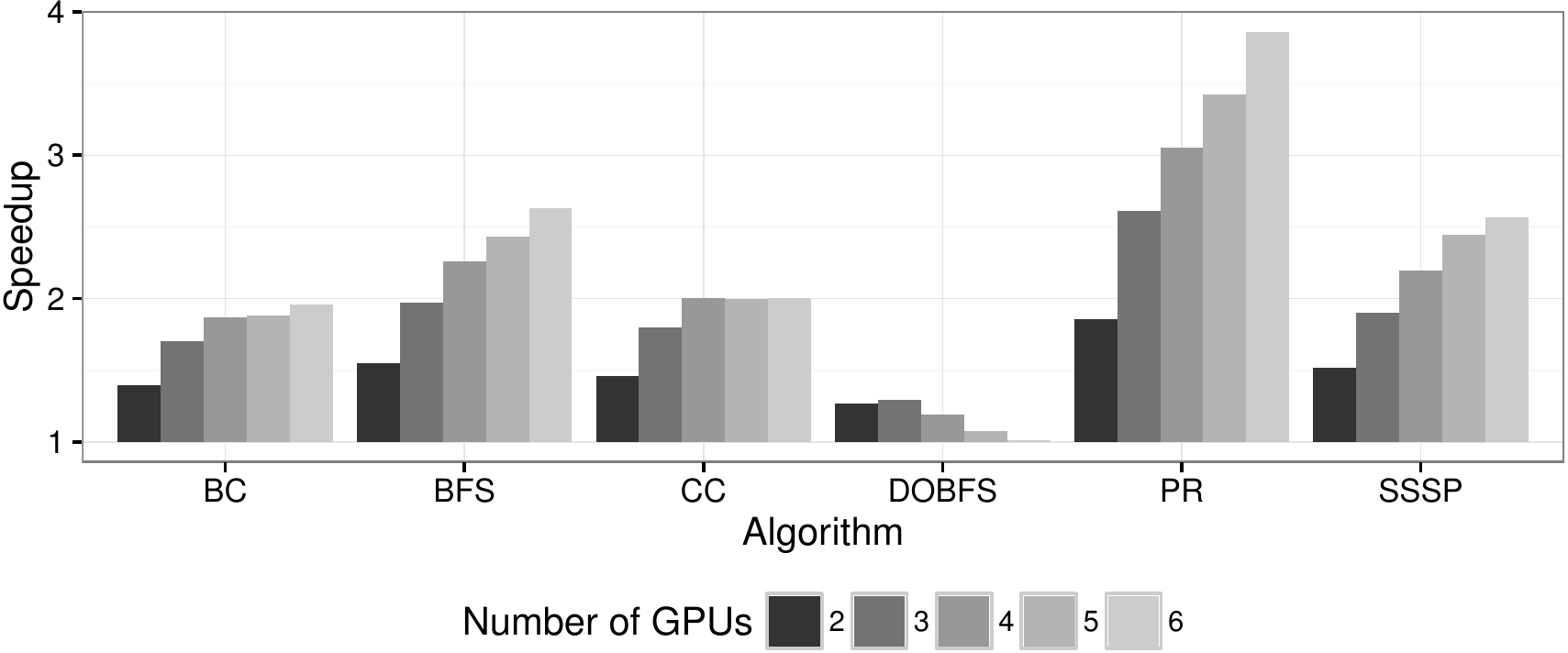}
  \centering
  \caption{mGPU speedup over 1GPU performance for BC, BFS, CC,
    DOBFS, PR and SSSP\@. All numbers shown are geometric means of
    runtime speedup over all datasets.\label{fig:overall} }
\end{figure}

\label{sec:dobfs-results}
The overall speedup of all the primitives is shown in
Fig.~\ref{fig:overall}, normalized to the performance of 1GPU
as 1. The speedup of a given primitive using a given number of GPUs is
the geometric mean of speedups from all datasets tested for that
configuration. Most of the primitives scale well from 1 to 6 GPUs,
resulting in 2.63$\times$, 2.57$\times$, 2.00$\times$, 1.96$\times$
and 3.86$\times$ speedup for BFS,
SSSP, CC, BC and PR respectively using 6 (K40) GPUs. The performance
curve of DOBFS mostly stays flat, as it's limited by communication
overhead. This agrees with our scaling analysis in
Section~\ref{sec:analysis} and ~\ref{sec:dobfs-optimization}.

\begin{figure*}
  \begin{minipage}[b]{0.34\linewidth}
    \includegraphics[width=\textwidth]{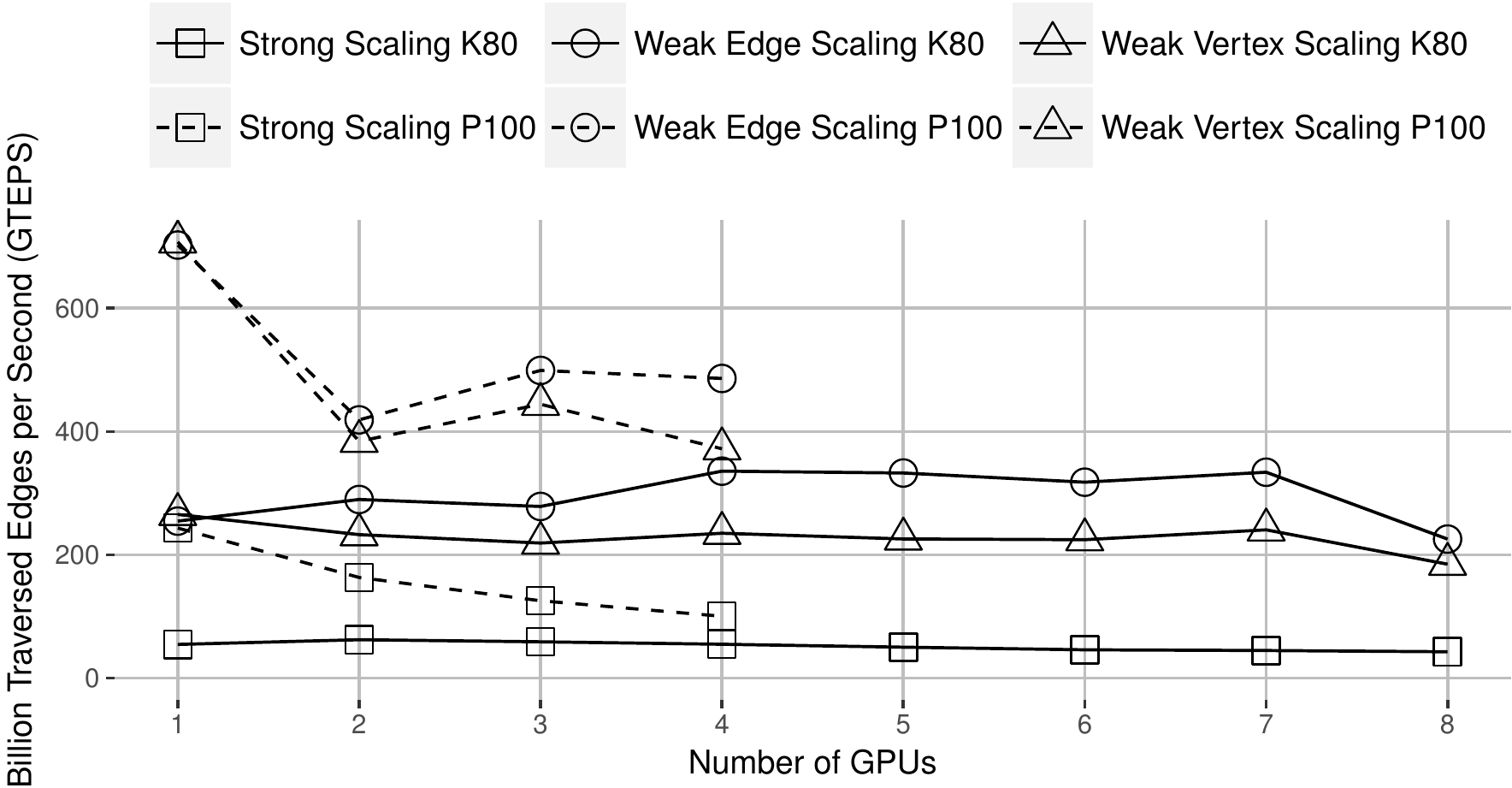}
    \subcaption{DOBFS scalability}
    \label{fig:DOBFS_Scaling}
  \end{minipage}
  \begin{minipage}[b]{0.34\linewidth}
    \includegraphics[width=\textwidth]{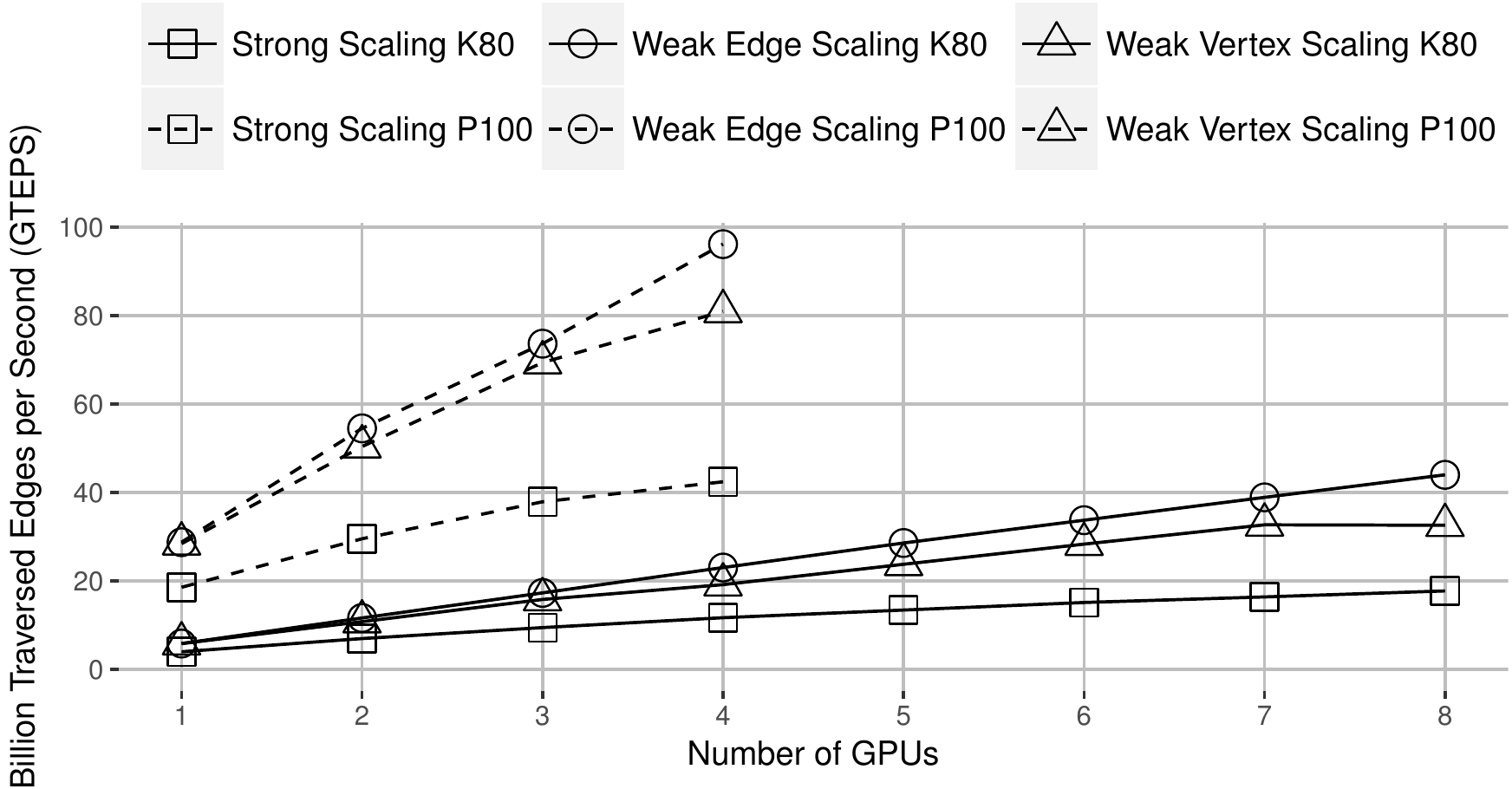}
    \subcaption{BFS scalability}
  \label{fig:BFS_Scaling}
  \end{minipage}
  \begin{minipage}[b]{0.34\linewidth}
    \includegraphics[width=\textwidth]{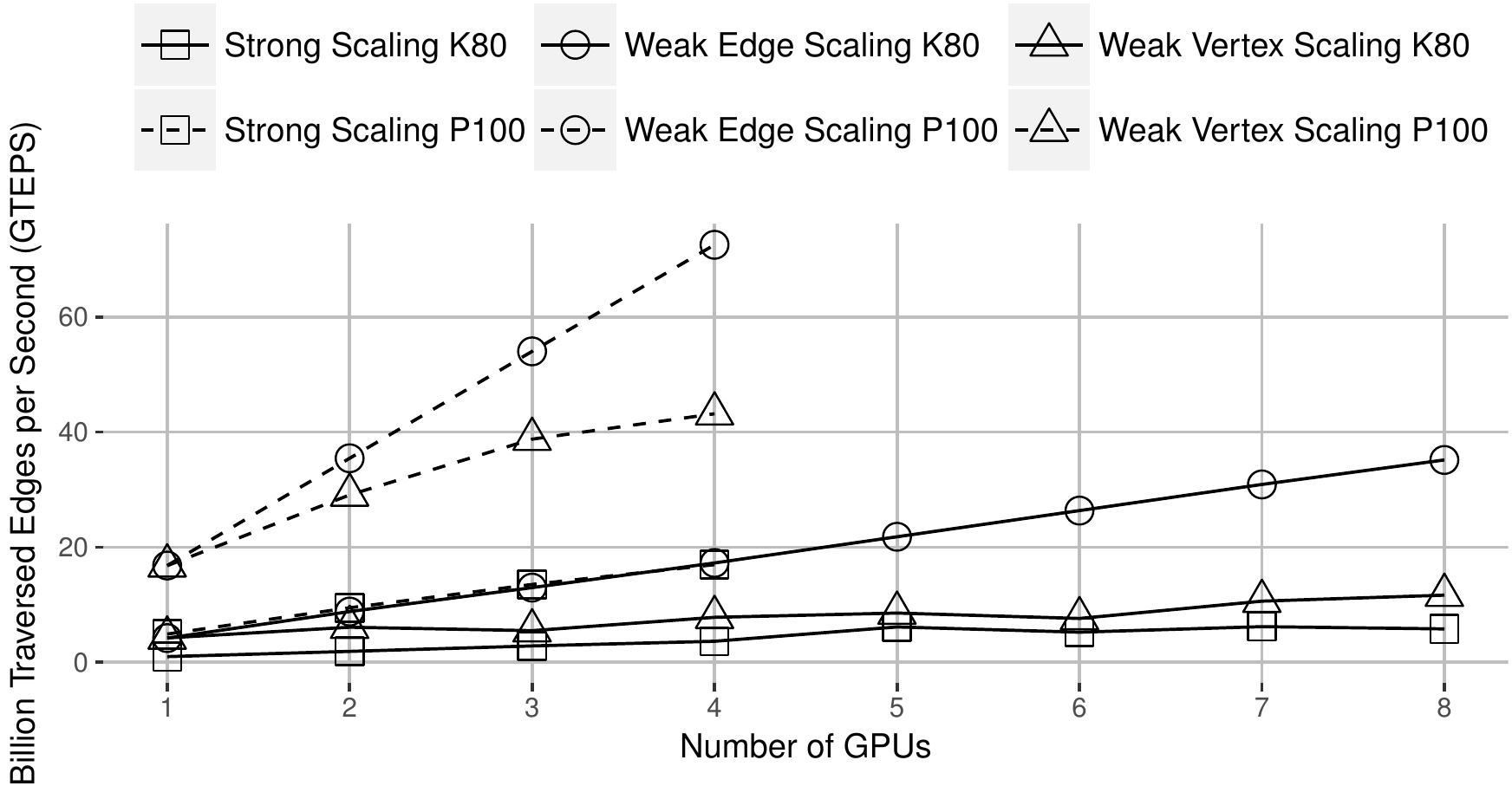}
    \subcaption{PR scalability}
  \label{fig:PR_Scaling}
  \end{minipage}
  \caption{Scalability of DOBFS, BFS, and PR\@. \{Strong, weak edge, weak
    vertex\} scaling use rmat graphs with \{$2^{24}$, $2^{19}$,
    $2^{19}\times|\text{GPUs}|$\} vertices and edge
    factor \{32, 256$\times|\text{GPUs}|$, 256\} respectively.
    %Strong scaling
    %experiments use rmat graphs with scale 24 and edge factor 32. Weak
    %edge scaling experiments use rmat graphs with scale 19 and edge
    %factor 256$\times|\text{GPUs}|$, while weak vertex scaling experiments
    %use rmat graphs with $2^{19}\times|\text{GPUs}|$ vertices and edge factor
    %256.
    \label{fig:DOBFS_BFS_PR_Scaling} }
\end{figure*}

As in Section~\ref{sec:algorithms}, we focus on (DO)BFS and PR as
representative primitives for further analysis.
Fig.~\ref{fig:DOBFS_BFS_PR_Scaling} shows strong and weak
scaling of these selected primitives. While providing both weak-vertex
and -edge scaling, DOBFS doesn't have good strong scaling, because
its computation and communication are both roughly in the order of
$O(|V_i|)$. This effect is more obvious on P100, as computation is
faster but inter-GPU bandwidth stays mostly the same. As a result, the
peak BFS performance (513~GTEPS on K40, and 900~GTEPS on P100) is
achieved by 1GPU DOBFS with rmat\_n20\_512. In
contrast, BFS and PR achieve almost linear weak and strong scaling
from 1 to 8 GPUs.

\begin{figure}
  \centering
  \includegraphics[width=\columnwidth]{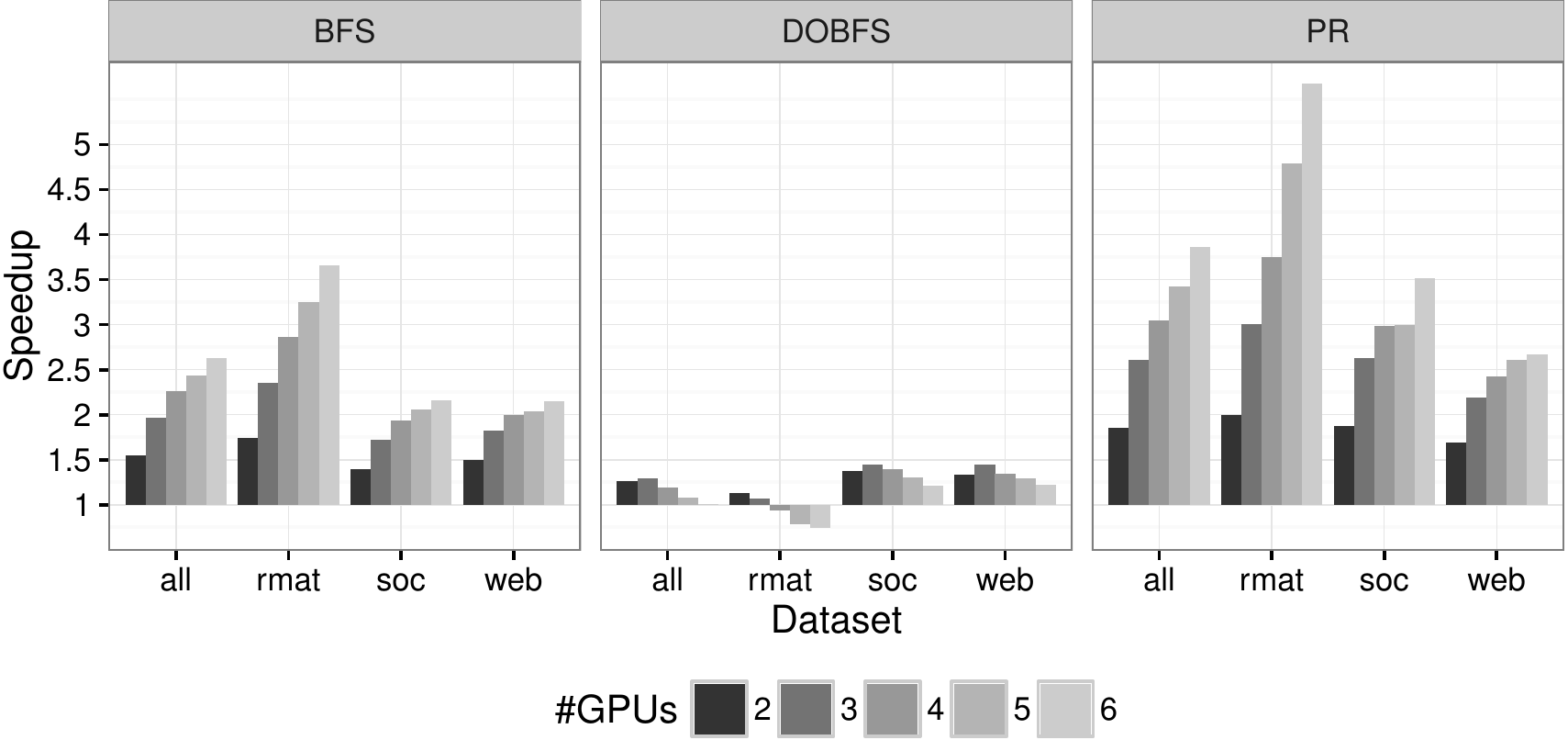}
  \centering
  \caption{mGPU geometric mean speedups over 1GPU performance on
    rmat, soc and web graphs separately and in geometric means (all) for
    DOBFS, BFS, and PR\@.\label{fig:speedup}}
\end{figure}

We show more detailed speedups separated by graph type in
Fig.~\ref{fig:speedup}. DOBFS scaling suffers the most for rmat
datasets, because the volume and computation for communication vs.\
the core computation complexity is higher, resulting in a
correspondingly larger portion of per-iteration time for inter-GPU
communications. On the other hand, the larger $|E_i| / |V_i|$ ratio of
rmat graphs helps BFS and PR in scalability, because the core
computation cost is $O(|E_i|)$ and the communication cost is at most
$O(|V_i|)$, reducing the cost of communication compared to
computation.

%% \input {table/road.tex}
% \begin{figure}
%  \centering
%  \includegraphics[width=\columnwidth]{fig/Road.pdf}
%  \centering
%  \caption{Geomean slowdown of road networks with more GPUs, measured on 5
%    DIMACS10 road networks~\cite{DIMACS:2011:1DI}  (germany\_osm, asia\_osm, road\_central,
%    road\_usa, europe\_osm).\label{fig:road}}
% \end{figure}

% \noindent
% \textbf{Road networks}
% \label{sec:road}
% Prior mGPU graph analytics have largely focused on large
% scale-free graphs with small diameters. These graphs are easier to
% parallelize and scale than high-diameter, low-degree graphs like road
% networks. In fact, scalability is a significant
% problem with these graphs on our system (Figure~\ref{fig:road}), and
% we expect that if other frameworks evaluated these graphs, they would
% see similar results. With our approach, these graphs have insufficient
% parallelism to saturate even a single GPU, much less multiple GPUs. As
% well, iteration overhead occupies a significant portion of the elapsed
% time. As an example, on germany\_osm, the iteration overhead alone is
% \{71.4\%, 55\%\} of the total runtime on \{1, 2\} GPUs. Efficiently
% running these graphs would require a different approach, perhaps
% requiring preprocessing~\cite{Delling:2010:PHS}.

\subsection{Comparisons vs.\ Previous mGPU Work}
\label{sec:compare}
\begin{table*}[t]
  \centering
    \resizebox{\textwidth}{!}{\begin{tabular}{@{}lcccccc@{}}
      \toprule
      \multicolumn{1}{c}{graph} %& algo
      & ref. & ref.\ hw. & ref.\ perf. & our hw. & our perf. & comp. \\
      \midrule

      kron\_n24\_32 (16.8M, 1.07B, UD) %& BFS
    & Liu~\cite{Liu:2015:EBG} & \{2, 4\}$\times$K40$\times$1 & \{15, 18\}~GTEPS
    & \{2, 4\}$\times$K40 & \{77.7, 67.7\}~GTEPS & \{5.18, 3.76\}$\times$ \\

      kron\_n24\_32 (16.8M, 1.07B UD) %& BFS
    & Liu~\cite{Liu:2015:EBG} & 8$\times$K40$\times$1 & 18.4~GTEPS
    & 4$\times$K80 & 40.2~GTEPS & 2.18$\times$ \\

      rmat\_2Mv\_128Me (2M, 128M, D) %& BFS
    & Merrill~\cite{Merrill:2012:SGG} & 4$\times$K40$\times$1$^N$ & 11.2~GTEPS
    & 4$\times$K40 & 29.9~GTEPS & 2.67$\times$ \\

      coPapersCiteseer (0.43M, 32.1M, UD) %& BFS
    & Zhong~\cite{Zhong:2014:MSG} & 4$\times$C2050$\times$1$^E$ & 2.69~GTEPS
    & 4$\times$K40 & 3.31~GTEPS & 1.23$\times$ \\

      com-orkut (3M, 117M, UD) %& BFS
    & Bisson~\cite{Bisson:2015:PDB} & 1$\times$K20X$\times$4& 2.67~GTEPS
    & 4$\times$K40 & 14.22~GTEPS & 5.33$\times$ $|$ 4.62$\times^*$ \\

      com-Friendster (66M, 1.81B, UD) %& BFS
    & Bisson~\cite{Bisson:2015:PDB} & 1$\times$K20X$\times$64& 15.68~GTEPS
    & 4$\times$K40 & 14.1~GTEPS & 0.90$\times$ $|$ 0.78$\times^*$ \\

      kron\_n23\_16 (8M, 256M, UD) %& BFS
    & Bernaschi~\cite{Bernaschi:2015:EGD} & 1$\times$K20X$\times$4& $\sim$1.3~GTEPS
    & 4$\times$K40 & 30.8~GTEPS & 23.7$\times$ $|$ 20.6$\times^*$ \\

      kron\_n25\_16 (32M, 1.07B, UD) %& BFS
    & Bernaschi~\cite{Bernaschi:2015:EGD} & 1$\times$K20X$\times$16& $\sim$3.2~GTEPS
    & 6$\times$K40 & 31.0~GTEPS & 9.69$\times$ $|$ 8.41$\times^*$ \\

      kron\_n25\_32 (32M, 1.07B, D) %& BFS
    & Fu~\cite{Fu:2014:PBF} & 2$\times$K20$\times$32& 22.7~GTEPS
    & 4$\times$K40 & 32.0~GTEPS & 1.41$\times$ $|$ 1.02$\times^*$ \\

      kron\_n23\_32 (8M, 256M, D) %& BFS
    & Fu~\cite{Fu:2014:PBF} & 2$\times$K20$\times$2& 6.3~GTEPS
    & 4$\times$K40 & 27.9~GTEPS & 4.43$\times$ $|$ 3.20$\times^*$ \\

      twitter-mpi (52.6M, 1.96B, D) %& BFS
    & Bebee~\cite{Bebee:2016:WTD} & 1$\times$K40$\times$16 & 224.2 ms
    & 3$\times$K40 & 94.31 ms & 2.38$\times$ \\

      \bottomrule
    \end{tabular}}
    \caption{Comparison with previous in-core GPU BFS work. Ref.\
      hardware is denoted by intra-node GPU count$\times$GPU
      model$\times$node count. We use the same number of GPUs whenever
      possible within the constraints of a single node. $*$ indicates
      speedup adjustment by memory bandwidth ratio, $N$ indicates
      results reproduced on our system, and $E$ indicates issues in
      reproducing results; refer to Section~\ref{sec:compare} for
      details.\label{tab:compare}}
\end{table*}

\newcommand{\specialcell}[2][c]{%
  \begin{tabular}[#1]{@{}c@{}}#2\end{tabular}}

\begin{table*}[t]
  \centering
  \resizebox{\textwidth}{!}{\begin{tabular}{@{}lccccccc@{}}
      \toprule
      \multicolumn{1}{c}{graph} & algo & ref. & ref.\ perf. & our hw. & our perf. \\
      \midrule
      uk-2002 & \{BFS, SSSP, CC, PR\}
    & Sengupta~\cite{Sengupta:2015:GPL}, 1$\times$K40$^E$
    & \{49, 80, 153, 162\} sec
    & 1$\times$K40 & \{0.059, 0.76, 1.85, 1.99\} sec \\

      twitter-rv & \{BFS, SSSP, CC, PR\}
    & Shi~\cite{Shi:2015:FAG}, 1$\times$K40$^N$ & \{46, 40$^\ddagger$, 29, 80\} sec
    & \{1, 2, 3, 1\}$\times$K40 & \{0.098, 0.837$^\ddagger$, 1.71, 49.7\} sec \\

      LiveJournal1 & \{BFS, SSSP, CC, PR\}
    & Shi~\cite{Shi:2015:OAG}, 1$\times$K40$^N$ & \{66.4, 245$^\ddagger$, 213, 105\} ms
    & 1$\times$K40 & \{12.2, 63.2$^\ddagger$, 93.6, 45.7\} ms \\

      twitter-rv & \{SSSP, CC, PR\}
    & Lee~\cite{Lee:2015:SIG}, 4 cores$\times$21 nodes
    & \{126, 304, 149\} sec
    & \{2, 3, 1\}$\times$K40 & \{2.20, 1.71, 49.7\} sec \\

      twitter-mpi & \{BFS, SSSP, BC, PR\}
    & \specialcell{Gharaibeh~\cite{Gharaibeh:2014:ELS}, 2$\times$\\K40+2$\times$Xeon
2637$^N$}
    & \specialcell{\{0.698, 2.67, 3.90, \\0.581 / iter\} sec}
    & 4$\times$K40 & \specialcell{\{0.0785, 1.62, 2.37, \\0.471 / iter\} sec} \\
      \bottomrule
    \end{tabular}}
    \caption{Comparison with previous out-of-core GPU or CPU graph
      processing work. Our framework can process the largest datasets
      that were reported by most previous works (except Totem), on all reported
      primitives, using much less processing time. \{uk-2002,
      twitter-rv, LiveJournal1, twitter-mpi\} are directed graphs with
      \{18.5M, 42M, 5M, 52.6M\} vertices and \{298M, 1.5B, 68M,
      1.96B\} edges. $N$ and $E$ indicators are the same as in
      Table~\ref{tab:compare}. $\ddagger$ notes that Frog uses 1 uniformly
      as edge weights for SSSP, which we also use for Frog+SSSP
      comparisons only.\label{tab:compare2}}
    \vspace{-1em}
\end{table*}

We compare our work with previous GPU in-core systems in
Table~\ref{tab:compare}, and with previous GPU out-of-core or CPU
systems in Table~\ref{tab:compare2}. The datasets we choose for
comparison against each system are those specifically highlighted by
the authors in their results, presumably the
datasets where their systems show the best results. We make our best
efforts to reproduce all reported results from open-source single-node
implementations on our system for direct comparison. Some of them run
into issues with reproducibility, and we have communicated with the
respective authors of the libraries to resolve these issues. Some
reported results use K20 GPUs; as we do not have access to this
particular GPU, for these comparisons, we instead scale our speedups
by the memory bandwidth ratio between the K20 and K40 we
use. This comparison disfavors Gunrock because we verified that
Gunrock's relative performance reduction is always smaller than the
relative memory bandwidth reduction on Kepler GPUs for large rmat and
social networks.

% While our primary comparisons are
% against what we feel is the fairest comparison---systems with the same
% number of GPUs---we also note that our system features all of those
% GPUs on the same node, whereas some of the comparison systems instead
% allocate one GPU per node.

% Finally, most of the systems that
% we compare against are designed to run only a single graph primitive,
% and are optimized specifically for that primitive, whereas our system
% runs a wide range of graph primitives in a programmable framework
% without a primitive-specific focus.

%\paragraph{Details of particular comparisons}

Enterprise~\cite{Liu:2015:EBG} is a hardwired DOBFS implementation
with various optimizations. It is considered state of the art for
a traditional DOBFS implementation on GPUs within a single node.
Our DOBFS outperforms it by 2--5$\times$,
even given less than ideal scalability with DOBFS and rmat. The
results of the BFS-specific implementation in B40C by Merrill et
al.~\cite{Merrill:2012:SGG} without directional optimization are
particularly impressive; to be consistent with other comparisons, our
29.9~GTEPS result is produced by DOBFS\@; our normal BFS
records 12.9~GTEPS, $\sim$1.15$\times$ compared to B40C\@. We use
Merrill's rmat parameters (\{A, B, C, D\} = \{0.45, 0.15, 0.15, 0.25\})
for this particular comparison.

The graphs selected by Zhong et al.~\cite{Zhong:2014:MSG} are not
considered as large ones, and most of our runtime would be on
iteration overhead introduced by Gunrock's load balancing steps that
are more useful for large graphs. Despite that, we still see 1.23$\times$
speedup as compared to their best BFS result.

Works by Bisson et al.~\cite{Bisson:2015:PDB}, Bernaschi et
al.~\cite{Bernaschi:2015:EGD}, Fu et al.~\cite{Fu:2014:PBF}, and Bebee
et al.~\cite{Bebee:2016:WTD} are GPU-cluster-based implementations.
Our results with 4--6 GPUs show significant speedup compared to theirs
with 4--16 GPUs in a cluster. Using 4 GPUs we achieve similar
performance as their 64-GPU clusters. We note that inter-GPU bandwidth
within a node is larger than inter-node bandwidth, so our comparisons
must be considered in this light; however, as we noted in Section
~\ref{sec:intro}, we believe that our results motivate a future focus on
scaling up (fewer but more powerful nodes, each with more GPUs) in
preference to scaling out (more nodes).

GraphReduce~\cite{Sengupta:2015:GPL} and
Frog (asynchronous)~\cite{Shi:2015:FAG, Shi:2015:OAG} are
out-of-core GPU approaches,
GraphMap~\cite{Lee:2015:SIG} targets CPU distributed-memory clusters, and
Totem~\cite{Gharaibeh:2014:ELS} is an heterogeneous CPU-GPU
approach. While out-of-core approaches have the promise to process
graphs much larger than in-core work such as ours, our framework can
comfortably process the largest graphs they used in any of their
results~\cite{Sengupta:2015:GPL, Shi:2015:FAG, Shi:2015:OAG,
  Lee:2015:SIG}. For these comparisons, we use the smallest number of
GPUs possible for individual comparisons, and achieve much less
processing time. For comparisons with Totem, we use the same number of
processors (4 GPUs vs.\ 2 CPUs + 2 GPUs), and achieve better
performance. GPU memory capacity is certainly an important concern,
but careful memory management
(Section~\ref{sec:careful-memory-management}) can allow even mGPUs
to run graphs of significant size directly
from GPU memory. When graphs can fit into GPU memory, in-core is more
preferable than out-of-core in view of performance.

We also compare our work (using a system with an Intel Xeon E3 1225 v3
CPU and a single NVIDIA Tesla K40c GPU) with Daga et
al.~\cite{Daga:2014:EBF}. On 8 of the 9 graphs they used (the wiki
graph is no longer available online), Gunrock shows 5 to 10$\times$
performance (TEPS) as compared to Hybrid++(CPU+dGPU) and about
3.5$\times$ efficiency (TEPS per Watt) as compared to Hybrid++(APU),
with the exception of the road network, in which Gunrock's performance
and efficiency are only half of Daga's. Although the APU provides the
GPU with direct access to the main memory, its overall limited
bandwidth bottlenecks its performance.

Compared to previous work, the performance advantages of our framework come from:
\begin{itemize}
\item our novel optimizations (Section~\ref{sec:optimization}) that
  speed up computation or reduce memory usage;

\item using \texttt{cudaStream} to asynchronously launch computation
  and communication workloads, and \texttt{cudaEvent} to establish
  workload dependencies, allowing overlapping workloads when possible;

\item additional computation required by the framework is as
  lightweight as possible, reducing mGPU overhead; and

\item using high-performance, extensible single-GPU primitives as our
  building blocks.
\end{itemize}

\subsection{Larger Graphs}
\begin{table} %[t]
  \centering
    \begin{tabular}{lcc}
      \toprule
      \multicolumn{1}{c}{graph} & algo & perf. \\
      \midrule

      friendster (125M, 3.62B, UD) & BFS & 339 ms \\
      friendster (125M, 2.59B, D)  & PR  & 1024 ms / iter \\
      sk-2005 (50.6M, 1.9B, D)     & BFS & 2717 ms \\
      sk-2005 (50.6M, 1.9B, D)     & PR  & 154 ms / iter \\
      rmat\_n24\_32 (UD, 32bit eID) & BFS & 67.6~GTEPS \\
      rmat\_n24\_32 (UD, 64bit eID) & BFS & 52.6~GTEPS \\
      rmat\_n24\_32 (UD, 64bit vID) & BFS & 33.9~GTEPS
\\
      \bottomrule
    \end{tabular}
  \caption{Our performance on large graphs.\label{tab:larger}}
\end{table}

We also ran tests on larger graphs on 4 GPUs (Table~\ref{tab:larger}).
We achieve good performance on graphs up to 3.62B edges. As graphs
approach larger sizes, 32-bit vertex and edge IDs are no longer
sufficient, so our system supports 64-bit vertex and edge IDs. In
practice this doubles bandwidth requirements and our performance drops
accordingly. For example, on \texttt{rmat\_n24\_32}, BFS with 64-bit vertex ID
reads 2$\times$ data per edge as 32-bit, and
records 0.5$\times$ performance.

\section{Conclusions}
\label{sec:conclusion}
Increasing graph sizes and performance requirements provided the motivation to explore graph analytics on multiple GPUs. The size concern is particularly pressing for the limited memory space in current GPUs. Our chief goals were generality (can target many graph algorithms), programmability (particularly a simple extension from single-GPU programs to the multi-GPU ones), and scalability in performance and memory usage.

The most helpful decision we made was our unified framework for authoring a range of graph primitives, with high-level programmability for expressing the primitives and common components to extend these primitives to multiple GPUs. One challenge was the design of our abstraction that allowed both multi-GPU generality/programmability and scalable performance, but doing so both allowed a straightforward extension for programmers from single to multiple GPUs, as well as a higher level view of the key building blocks of a multi-GPU implementation, showing which operations are common to multiple algorithms, and what optimizations can be done at the framework level. As a result, improvements we make to the core of our framework apply to all graph primitives.

We see two key next steps. First, while we achieve good scalability in most cases, road networks and DOBFS do not scale well. How can we tackle these graphs from a systems perspective, whether that be GPU/platform hardware, system software, or our platform software? Second, can we achieve further scalability (scale-out) with multiple nodes, and given the increased latency and decreased bandwidth of those nodes, is it profitable to do so?

\section*{Acknowledgments}
\label{sec:ack}
Thanks to our DARPA program managers Wade Shen and Christopher White,
and DARPA business manager Gabriela Araujo, for their support during
this project. We appreciate the technical assistance, advice, and
machine access from many colleagues at NVIDIA\@: Chandra Cheij, Joe
Eaton, Michael Garland, Mark Harris, Duane Merrill, and Nikolai
Sakharnykh. Thanks also to our colleagues at Onu Technology: Erich
Elsen, Guha Jayachandran, and Vishal Vaidyanathan. Thanks to Roger
Pearce and Ayd\i{}n Bulu\c{c} for helpful comments along the way.

We gratefully acknowledge the support of the DARPA XDATA program (US
Army award W911QX-12-C-0059); DARPA STTR awards D14PC00023 and
D15PC00010; and NSF awards CCF-1017399, OCI-1032859, and CCF-1629657.

\bibliographystyle{IEEEtran}
\bibliography{bib/all,refs,temp}

\begin{thebibliography}{10}
\providecommand{\url}[1]{#1}
\csname url@samestyle\endcsname
\providecommand{\newblock}{\relax}
\providecommand{\bibinfo}[2]{#2}
\providecommand{\BIBentrySTDinterwordspacing}{\spaceskip=0pt\relax}
\providecommand{\BIBentryALTinterwordstretchfactor}{4}
\providecommand{\BIBentryALTinterwordspacing}{\spaceskip=\fontdimen2\font plus
\BIBentryALTinterwordstretchfactor\fontdimen3\font minus
  \fontdimen4\font\relax}
\providecommand{\BIBforeignlanguage}[2]{{%
\expandafter\ifx\csname l@#1\endcsname\relax
\typeout{** WARNING: IEEEtran.bst: No hyphenation pattern has been}%
\typeout{** loaded for the language `#1'. Using the pattern for}%
\typeout{** the default language instead.}%
\else
\language=\csname l@#1\endcsname
\fi
#2}}
\providecommand{\BIBdecl}{\relax}
\BIBdecl

\bibitem{Keckler:2011:GAT}
S.~W. Keckler, W.~J. Dally, B.~Khailany, M.~Garland, and D.~Glasco, ``{GPU}s
  and the future of parallel computing,'' \emph{IEEE Micro}, vol.~31, no.~5,
  pp. 7--17, Sep. 2011.

\bibitem{Zhong:2014:MSG}
J.~Zhong and B.~He, ``Medusa: Simplified graph processing on {GPU}s,''
  \emph{IEEE Transactions on Parallel and Distributed Systems}, vol.~25, no.~6,
  pp. 1543--1552, Jun. 2014.

\bibitem{Fu:2014:MAH}
Z.~Fu, M.~Personick, and B.~Thompson, ``{MapGraph}: A high level {API} for fast
  development of high performance graph analytics on {GPU}s,'' in
  \emph{Proceedings of the Workshop on GRAph Data Management Experiences and
  Systems}, ser. GRADES '14, Jun. 2014, pp. 2:1--2:6.

\bibitem{Khorasani:2014:CVG}
F.~Khorasani, K.~Vora, R.~Gupta, and L.~N. Bhuyan, ``{CuSha}: Vertex-centric
  graph processing on {GPUs},'' in \emph{Proceedings of the 23rd International
  Symposium on High-performance Parallel and Distributed Computing}, ser. HPDC
  '14, Jun. 2014, pp. 239--252.

\bibitem{Wang:2016:GAH:nourl}
Y.~Wang, A.~Davidson, Y.~Pan, Y.~Wu, A.~Riffel, and J.~D. Owens, ``{G}unrock: A
  high-performance graph processing library on the {GPU},'' in
  \emph{Proceedings of the 21st ACM SIGPLAN Symposium on Principles and
  Practice of Parallel Programming}, ser. PPoPP 2016, Mar. 2016.

\bibitem{Beamer:2012:DBS}
S.~Beamer, K.~Asanovi\'{c}, and D.~Patterson, ``Direction-optimizing
  breadth-first search,'' in \emph{Proceedings of the International Conference
  on High Performance Computing, Networking, Storage and Analysis}, ser. SC
  '12, Nov. 2012, pp. 12:1--12:10.

\bibitem{Merrill:2012:SGG}
D.~Merrill, M.~Garland, and A.~Grimshaw, ``Scalable {GPU} graph traversal,'' in
  \emph{Proceedings of the 17th ACM SIGPLAN Symposium on Principles and
  Practice of Parallel Programming}, ser. PPoPP '12, Feb. 2012, pp. 117--128.

\bibitem{Bisson:2015:PDB}
M.~Bisson, M.~Bernaschi, and E.~Mastrostefano, ``Parallel distributed breadth
  first search on the {K}epler architecture,'' \emph{IEEE Transactions on
  Parallel and Distributed Systems}, vol.~PP, no.~99, Sep. 2015.

\bibitem{Liu:2015:EBG}
H.~Liu and H.~H. Huang, ``Enterprise: Breadth-first graph traversal on
  {GPU}s,'' in \emph{Proceedings of the International Conference for High
  Performance Computing, Networking, Storage and Analysis}, ser. SC '15, Nov.
  2015, pp. 68:1--68:12.

\bibitem{McLaughlin:2014:SAH}
A.~McLaughlin and D.~A. Bader, ``Scalable and high performance betweenness
  centrality on the {GPU},'' in \emph{Proceedings of the International
  Conference for High Performance Computing, Networking, Storage and Analysis},
  ser. SC14, Nov. 2014, pp. 572--583.

\bibitem{Karypis:1998:FHQ}
G.~Karypis and V.~Kumar, ``A fast and high quality multilevel scheme for
  partitioning irregular graphs,'' \emph{SIAM J. Sci. Comput.}, vol.~20, no.~1,
  pp. 359--392, Dec. 1998.

\bibitem{Soman:2010:AFG}
J.~Soman, K.~Kishore, and P.~J. Narayanan, ``A fast {GPU} algorithm for graph
  connectivity,'' in \emph{24th IEEE International Symposium on Parallel and
  Distributed Processing, Workshops and PhD Forum}, ser. IPDPSW 2010, Apr.
  2010, pp. 1--8.

\bibitem{Gharaibeh:2014:ELS}
A.~Gharaibeh, T.~Reza, E.~Santos-Neto, L.~B. Costa, S.~Sallinen, and
  M.~Ripeanu, ``Efficient large-scale graph processing on hybrid {CPU} and
  {GPU} systems,'' \emph{CoRR}, vol. abs/1312.3018, no. 1312.3018v2, Dec. 2014.

\bibitem{Daga:2014:EBF}
M.~Daga, M.~Nutter, and M.~Meswani, ``Efficient breadth-first search on a
  heterogeneous processor,'' in \emph{IEEE International Conference on Big
  Data}, Oct. 2014, pp. 373--382.

\bibitem{Sengupta:2015:GPL}
D.~Sengupta, S.~L. Song, K.~Agarwal, and K.~Schwan, ``{GraphReduce}: Processing
  large-scale graphs on accelerator-based systems,'' in \emph{Proceedings of
  the International Conference for High Performance Computing, Networking,
  Storage and Analysis}, Nov. 2015, pp. 28:1--28:12.

\bibitem{Shi:2015:OAG}
X.~Shi, J.~Liang, S.~Di, B.~He, H.~Jin, L.~Lu, Z.~Wang, X.~Luo, and J.~Zhong,
  ``Optimization of asynchronous graph processing on {GPU} with hybrid coloring
  model,'' in \emph{Proceedings of the 20th ACM SIGPLAN Symposium on Principles
  and Practice of Parallel Programming}, ser. PPoPP 2015, Feb. 2015, pp.
  271--272.

\bibitem{Shi:2015:FAG}
X.~Shi, X.~Luo, J.~Liang, P.~Zhao, S.~Di, B.~He, and H.~Jin, ``Frog:
  Asynchronous graph processing on {GPU} with hybrid coloring model,'' Huazhong
  University of Science and Technology, Tech. Rep. HUST-CGCL-TR-402, 2015,
  \url{http://grid.hust.edu.cn/xhshi/projects/frog.html}.

\bibitem{Ben-Nun:2017:GAA}
T.~Ben-Nun, M.~Sutton, S.~Pai, and K.~Pingali, ``Groute: An asynchronous
  multi-{GPU} programming model for irregular computations,'' in
  \emph{Proceedings of the 22nd ACM SIGPLAN Symposium on Principles and
  Practice of Parallel Programming}, ser. PPoPP 2017, Feb. 2017.

\bibitem{Valiant:1990:ABM}
L.~G. Valiant, ``A bridging model for parallel computation,''
  \emph{Communications of the ACM}, vol.~33, no.~8, pp. 103--111, 1990.

\bibitem{Wu:2015:PCF:nourl}
Y.~Wu, Y.~Wang, Y.~Pan, C.~Yang, and J.~D. Owens, ``Performance
  characterization for high-level programming models for {GPU} graph
  analytics,'' in \emph{IEEE International Symposium on Workload
  Characterization}, ser. IISWC-2015, Oct. 2015, pp. 66--75.

\bibitem{Davis:1994:UOF}
T.~A. Davis, ``The {U}niversity of {F}lorida sparse matrix collection,''
  \emph{NA Digest}, vol.~92, no.~42, 16~Oct. 1994,
  \url{http://www.cise.ufl.edu/research/sparse/matrices}.

\bibitem{Rossi:2015:NDR}
\BIBentryALTinterwordspacing
R.~A. Rossi and N.~K. Ahmed, ``The network data repository with interactive
  graph analytics and visualization,'' in \emph{Proceedings of the Twenty-Ninth
  AAAI Conference on Artificial Intelligence}, Mar. 2015, pp. 4292--4293.
  [Online]. Available: \url{http://networkrepository.com/}
\BIBentrySTDinterwordspacing

\bibitem{Bader:2006:GAS}
D.~A. Bader and K.~Madduri, ``{GT}graph: A suite of synthetic graph
  generators,'' 2006, \url{https://github.com/dhruvbird/GTgraph}.

\bibitem{Bernaschi:2015:EGD}
M.~Bernaschi, G.~Carbone, E.~Mastrostefano, M.~Bisson, and M.~Fatica,
  ``Enhanced {GPU}-based distributed breadth first search,'' in
  \emph{Proceedings of the 12th ACM International Conference on Computing
  Frontiers}, ser. CF '15, 2015, pp. 10:1--10:8.

\bibitem{Fu:2014:PBF}
Z.~Fu, H.~K. Dasari, B.~Bebee, M.~Berzins, and B.~Thompson, ``Parallel breadth
  first search on {GPU} clusters,'' in \emph{IEEE International Conference on
  Big Data}, Oct. 2014, pp. 110--118.

\bibitem{Bebee:2016:WTD}
B.~Bebee, ``What to do with all that bandwidth? {GPU}s for graph and predictive
  analytics,'' 21~Mar. 2016,
  \url{https://devblogs.nvidia.com/parallelforall/gpus-graph-predictive-analytics/}.

\bibitem{Lee:2015:SIG}
K.~Lee, L.~Liu, K.~Schwan, C.~Pu, Q.~Zhang, Y.~Zhou, E.~Yigitoglu, and P.~Yuan,
  ``Scaling iterative graph computations with {G}raph{M}ap,'' in
  \emph{Proceedings of the International Conference for High Performance
  Computing, Networking, Storage and Analysis}, ser. SC '15, 2015, pp.
  57:1--57:12.

\end{thebibliography}

\begin{appendices}
\section{Multi-GPU BFS code example}
\label{app:bfs}
This code list shows a mGPU BFS implementation using the
proposed framework. \textbf{Programmer provided, primitive
specific code} is highlighted. This implementation may not
cover all available optimizations, as the main purpose is
to illustrate how to extend a single GPU primitive onto mGPU.\\

\verbatimfont{\footnotesize}%
%\begin{lstlisting}[language=C++, basicstyle=\footnotesize]
\begin{Verbatim}[commandchars=\\\#\@, fontsize=\footnotesize]
struct BFSProblem : public ProblemBase {
// maximum number of associative values to
// send per-vertex, of type VertexT
static const int MAX_NUM_VERTEX_ASSOCIATES =
  \boldfont#(MARK_PREDECESSORS) ? 1 : 0@ ;
// maximum number of associative values to send
// send per-vertex, of types ValueT
static const int MAX_NUM_VALUE__ASSOCIATES = \boldfont#0@;

// Per-GPU problem specific data structure
struct DataSlice : BaseDataSlice {
  \boldfont#The same as on single GPU;@ };

Array1D<DataSlice> *data_slices;
BFSProblem() : BaseProblem(...) {}

void Init(Csr *graph, int num_gpus, ...) {
  // Init BaseProblem, include partitioning
  // of the graph, and generating
  // partition_tables and conversion_tables
  BaseProblem::Init(graph, num_gpus, ...);
  data_slices = new Array1D<DataSlice>[num_gpus];
  for (int gpu = 0; gpu < num_gpus; gpu++) {
    DataSlice &data_slice = data_slices[gpu];
    data_slice.Allocate(1, DEVICE | HOST);
    data_slice.Init(sub_graphs[gpu], ...);
    \boldfont#if (MARK_PREDECESSORS && num_gpus > 1)@
      \boldfont#data_slice.vertex_associate_orgs[0] =@
        \boldfont#preds.GetPointer(DEVICE);@
  }
}

// Reset data to be ready for new traversal
void Reset(VertexT src, ...) {
  for (int gpu = 0; gpu < num_gpus; gpu++)
    data_slices[gpu] -> Reset(...);
  // host GPU of the source vertex
  \boldfont#int src_gpu = 0;@
  // the Vertex Id of src on its host GPU
  \boldfont#VertexT tsrc = src;@
  \boldfont#if (num_gpus > 1) {@
    \boldfont#src_gpu = partition_tables[0][src];@
    \boldfont#tsrc = conertion_tables[0][src];@
  \boldfont#}@
  \boldfont#Init label and pred for tsrc on GPU src_gpu;@
  \boldfont#Put tsrc into initial frontier on GPU src_gpu;@
}
}; // end of struct BFSProblem

// Kernel to combine received and local data
__global__ void Expand_Incoming_Kernel(...) {
  SizeT i = blockIdx.x*blockDim.x + threadIdx.x;
  while (i < num_received_vertices) {
    VertexT v = received_vertices[i];
    \boldfont#if (label < atomicMin(@
      \boldfont#data_slice -> labels + key, label)) {@
      \boldfont#vertices_out[atomicAdd(out_length, 1)] = v;@
      \boldfont#if (MARK_PREDECESSORS)@
        \boldfont#data_slice -> preds[v] = vertex_associate_in[i];@
    \boldfont#}@
    i += blockDim.x * gridDim.x;
  }
}

struct BFSIteration : public IterationBase {
  static void Expand_Incoming(...) {
    Expand_Incoming_Kernel<<<...>>>(...); }

  // Core of BFS implementation, for 1 iteration
  static void FullQueue_Core(...) {
    \boldfont#Same as on single GPU;@}

  // BFS uses the default Stop_Condition(),
  // which exits the iteration loop when all
  // frontiers are empty, or any error occurs
};

// Control thread on CPU
... BFSThread(Thread_Slice *thread_slice) {
  thread_slice -> status = Idle;
  while (thread_slice -> status != ToKill) {
    while (thread_slice -> status == Wait ||
      thread_slice -> status == Idle)
        sleep(0);
    if (thread_slice -> status == ToKill) break;
    // Perform one BFS iteration loop
    gunrock::app::Iteration_Loop<BFSIteration>
      (thread_slice);
    thread_slice -> status = Idle;
  }
}

struct BFSEnactor : public EnactorBase {
  ThreadSlice *thread_slices;
  CUThread    *thread_ids;
  BFSEnactor(...) : EnactorBase(...), ... {}

  void Init(...) {
    BaseEnactor::Init(...);
    for (int gpu = 0; gpu < num_gpus; gpu++) {
      prepare thread_slices[gpu];
      thread_ids[gpu] = cutStartThread(
        BFSThread<...>, thread_slices[gpu]);
    }
    wait for all threads to be idle;
  }

  void Reset() {
    BaseEnactor::Reset();
    for (int gpu = 0; gpu < num_gpus; gpu++)
      thread_slices[gpu].status = Wait;
  }

  void Enact(VertexT src, ...) {
    Set initial frontier size on each GPU;
    //Signal GPUs to start working
    for (int gpu = 0; gpu < num_gpus; gpu++)
      thread_slices[gpu].status = Running;
    //Wait for GPUs to finish
    for (int gpu = 0; gpu < num_gpus; gpu++)
      while (thread_slices[gpu].status != Idle)
        sleep(0);
  }
}; // end of struct BFSEnactor

void BFS(Csr graph, int num_gpus, vector<> srcs) {
  BFSProblem problem();
  BFSEnactor enactor();
  problem.Init(graph, num_gpus, ...);
  enactor.Init(...);
  for (auto src : srcs) {
    problem.Reset(src, ...);
    enactor.Reset();
    // the actual traversal
    enactor.Enact(src, ...);
  }
}
\end{Verbatim}
%\end{lstlisting}

\end{appendices}
\end{document}